\documentclass[prd,twocolumn]{revtex4}
\usepackage[]{graphicx} 
\usepackage[]{units} 
\usepackage[]{color} 
\usepackage[]{amssymb} 
\usepackage[]{amsmath} 
\usepackage[]{hyperref} 
\usepackage[]{breakurl}
\usepackage[caption=false]{subfig} 
\usepackage[UKenglish]{babel}
\usepackage{multirow}
\newcommand{\nicesum}[0]{\displaystyle\sum}

\newcommand{\SfiveLivetime}{283.0 days}
\newcommand{\SsixLivetime}{132.9 days}
\newcommand{\SfiveTSLivetime}{84.1 years}
\newcommand{\SsixTSLivetime}{38.7 years}
\newcommand{\msuncd}{$\mathrm{M_{\odot} c^2}$}
\begin{document}


\title{All-sky search for long-duration gravitational wave transients with initial LIGO}


\author{%
B.~P.~Abbott$^{1}$,  
R.~Abbott$^{1}$,  
T.~D.~Abbott$^{2}$,  
M.~R.~Abernathy$^{1}$,  
F.~Acernese$^{3,4}$,
K.~Ackley$^{5}$,  
C.~Adams$^{6}$,  
T.~Adams$^{7}$,
P.~Addesso$^{8}$,  
R.~X.~Adhikari$^{1}$
% ...
\\
{(LIGO Scientific Collaboration and Virgo Collaboration)}%
}%
\email[]{publication@ligo.org; publication@ego-gw.it}
\medskip
\address {$^{\dag}$Deceased, May 2015. $^{\ddag}$Deceased, March 2015.}%
\medskip
\address {$^{1}$LIGO, California Institute of Technology, Pasadena, CA 91125, USA }
\address {$^{2}$Louisiana State University, Baton Rouge, LA 70803, USA }
\address {$^{3}$Universit\`a di Salerno, Fisciano, I-84084 Salerno, Italy }
\address {$^{4}$INFN, Sezione di Napoli, Complesso Universitario di Monte S.Angelo, I-80126 Napoli, Italy }
\address {$^{5}$University of Florida, Gainesville, FL 32611, USA }
\address {$^{6}$LIGO Livingston Observatory, Livingston, LA 70754, USA }
\address {$^{7}$Laboratoire d'Annecy-le-Vieux de Physique des Particules (LAPP), Universit\'e Savoie Mont Blanc, CNRS/IN2P3, F-74941 Annecy-le-Vieux, France }
\address {$^{8}$University of Sannio at Benevento, I-82100 Benevento, Italy and INFN, Sezione di Napoli, I-80100 Napoli, Italy }
\address {$^{9}$Albert-Einstein-Institut, Max-Planck-Institut f\"ur Gravi\-ta\-tions\-physik, D-30167 Hannover, Germany }
\address {$^{10}$Nikhef, Science Park, 1098 XG Amsterdam, The Netherlands }
\address {$^{11}$LIGO, Massachusetts Institute of Technology, Cambridge, MA 02139, USA }
\address {$^{12}$Instituto Nacional de Pesquisas Espaciais, 12227-010 S\~{a}o Jos\'{e} dos Campos, SP, Brazil }
\address {$^{13}$Inter-University Centre for Astronomy and Astrophysics, Pune 411007, India }
\address {$^{14}$International Centre for Theoretical Sciences, Tata Institute of Fundamental Research, Bangalore 560012, India }
\address {$^{15}$University of Wisconsin-Milwaukee, Milwaukee, WI 53201, USA }
\address {$^{16}$Leibniz Universit\"at Hannover, D-30167 Hannover, Germany }
\address {$^{17}$Universit\`a di Pisa, I-56127 Pisa, Italy }
\address {$^{18}$INFN, Sezione di Pisa, I-56127 Pisa, Italy }
\address {$^{19}$The University of Mississippi, University, MS 38677, USA }
\address {$^{20}$California State University Fullerton, Fullerton, CA 92831, USA }
\address {$^{21}$LAL, Univ Paris-Sud, CNRS/IN2P3, Orsay, France }
\address {$^{22}$Chennai Mathematical Institute, Chennai, India }
\address {$^{23}$University of Southampton, Southampton SO17 1BJ, United Kingdom }
\address {$^{24}$Universit\"at Hamburg, D-22761 Hamburg, Germany }
\address {$^{25}$INFN, Sezione di Roma, I-00185 Roma, Italy }
\address {$^{26}$Albert-Einstein-Institut, Max-Planck-Institut f\"ur Gravitations\-physik, D-14476 Potsdam-Golm, Germany }
\address {$^{27}$Montana State University, Bozeman, MT 59717, USA }
\address {$^{28}$Universit\`a di Perugia, I-06123 Perugia, Italy }
\address {$^{29}$INFN, Sezione di Perugia, I-06123 Perugia, Italy }
\address {$^{30}$European Gravitational Observatory (EGO), I-56021 Cascina, Pisa, Italy }
\address {$^{31}$Syracuse University, Syracuse, NY 13244, USA }
\address {$^{32}$SUPA, University of Glasgow, Glasgow G12 8QQ, United Kingdom }
\address {$^{33}$LIGO Hanford Observatory, Richland, WA 99352, USA }
\address {$^{34}$APC, AstroParticule et Cosmologie, Universit\'e Paris Diderot, CNRS/IN2P3, CEA/Irfu, Observatoire de Paris, Sorbonne Paris Cit\'e, F-75205 Paris Cedex 13, France }
\address {$^{35}$Wigner RCP, RMKI, H-1121 Budapest, Konkoly Thege Mikl\'os \'ut 29-33, Hungary }
\address {$^{36}$Columbia University, New York, NY 10027, USA }
\address {$^{37}$Stanford University, Stanford, CA 94305, USA }
\address {$^{38}$Universit\`a di Padova, Dipartimento di Fisica e Astronomia, I-35131 Padova, Italy }
\address {$^{39}$INFN, Sezione di Padova, I-35131 Padova, Italy }
\address {$^{40}$CAMK-PAN, 00-716 Warsaw, Poland }
\address {$^{41}$Astronomical Observatory Warsaw University, 00-478 Warsaw, Poland }
\address {$^{42}$University of Birmingham, Birmingham B15 2TT, United Kingdom }
\address {$^{43}$Universit\`a degli Studi di Genova, I-16146 Genova, Italy }
\address {$^{44}$INFN, Sezione di Genova, I-16146 Genova, Italy }
\address {$^{45}$RRCAT, Indore MP 452013, India }
\address {$^{46}$Faculty of Physics, Lomonosov Moscow State University, Moscow 119991, Russia }
\address {$^{47}$SUPA, University of the West of Scotland, Paisley PA1 2BE, United Kingdom }
\address {$^{48}$University of Western Australia, Crawley, Western Australia 6009, Australia }
\address {$^{49}$Department of Astrophysics/IMAPP, Radboud University Nijmegen, P.O. Box 9010, 6500 GL Nijmegen, The Netherlands }
\address {$^{50}$ARTEMIS, Universit\'e C\^ote d'Azur, CNRS and Observatoire de la C\^ote d'Azur, F-06304 Nice, France }
\address {$^{51}$MTA E\"otv\"os University, ``Lendulet'' Astrophysics Research Group, Budapest 1117, Hungary }
\address {$^{52}$Institut de Physique de Rennes, CNRS, Universit\'e de Rennes 1, F-35042 Rennes, France }
\address {$^{53}$Washington State University, Pullman, WA 99164, USA }
\address {$^{54}$Universit\`a degli Studi di Urbino 'Carlo Bo', I-61029 Urbino, Italy }
\address {$^{55}$INFN, Sezione di Firenze, I-50019 Sesto Fiorentino, Firenze, Italy }
\address {$^{56}$University of Oregon, Eugene, OR 97403, USA }
\address {$^{57}$Laboratoire Kastler Brossel, UPMC-Sorbonne Universit\'es, CNRS, ENS-PSL Research University, Coll\`ege de France, F-75005 Paris, France }
\address {$^{58}$VU University Amsterdam, 1081 HV Amsterdam, The Netherlands }
\address {$^{59}$University of Maryland, College Park, MD 20742, USA }
\address {$^{60}$Center for Relativistic Astrophysics and School of Physics, Georgia Institute of Technology, Atlanta, GA 30332, USA }
\address {$^{61}$Laboratoire des Mat\'eriaux Avanc\'es (LMA), IN2P3/CNRS, Universit\'e de Lyon, F-69622 Villeurbanne, Lyon, France }
\address {$^{62}$Universitat de les Illes Balears---IEEC, E-07122 Palma de Mallorca, Spain }
\address {$^{63}$Universit\`a di Napoli 'Federico II', Complesso Universitario di Monte S.Angelo, I-80126 Napoli, Italy }
\address {$^{64}$NASA/Goddard Space Flight Center, Greenbelt, MD 20771, USA }
\address {$^{65}$Canadian Institute for Theoretical Astrophysics, University of Toronto, Toronto, Ontario M5S 3H8, Canada }
\address {$^{66}$Tsinghua University, Beijing 100084, China }
\address {$^{67}$University of Michigan, Ann Arbor, MI 48109, USA }
\address {$^{68}$Universit\`a di Roma Tor Vergata, I-00133 Roma, Italy }
\address {$^{69}$INFN, Sezione di Roma Tor Vergata, I-00133 Roma, Italy }
\address {$^{70}$National Tsing Hua University, Hsinchu City, Taiwan 30013, R.O.C. }
\address {$^{71}$Charles Sturt University, Wagga Wagga, New South Wales 2678, Australia }
\address {$^{72}$University of Chicago, Chicago, IL 60637, USA }
\address {$^{73}$Caltech CaRT, Pasadena, CA 91125, USA }
\address {$^{74}$Korea Institute of Science and Technology Information, Daejeon 305-806, Korea }
\address {$^{75}$Australian National University, Canberra, Australian Capital Territory 0200, Australia }
\address {$^{76}$Carleton College, Northfield, MN 55057, USA }
\address {$^{77}$INFN, Gran Sasso Science Institute, I-67100 L'Aquila, Italy }
\address {$^{78}$Universit\`a di Roma 'La Sapienza', I-00185 Roma, Italy }
\address {$^{79}$University of Brussels, Brussels 1050, Belgium }
\address {$^{80}$Texas Tech University, Lubbock, TX 79409, USA }
\address {$^{81}$Cardiff University, Cardiff CF24 3AA, United Kingdom }
\address {$^{82}$University of Minnesota, Minneapolis, MN 55455, USA }
\address {$^{83}$The University of Melbourne, Parkville, Victoria 3010, Australia }
\address {$^{84}$The University of Texas Rio Grande Valley, Brownsville, TX 78520, USA }
\address {$^{85}$The University of Sheffield, Sheffield S10 2TN, United Kingdom }
\address {$^{86}$Montclair State University, Montclair, NJ 07043, USA }
\address {$^{87}$Universit\`a di Trento, Dipartimento di Fisica, I-38123 Povo, Trento, Italy }
\address {$^{88}$INFN, Trento Institute for Fundamental Physics and Applications, I-38123 Povo, Trento, Italy }
\address {$^{89}$The Pennsylvania State University, University Park, PA 16802, USA }
\address {$^{90}$University of Cambridge, Cambridge CB2 1TN, United Kingdom }
\address {$^{91}$Indian Institute of Technology, Gandhinagar Ahmedabad Gujarat 382424, India }
\address {$^{92}$Institute for Plasma Research, Bhat, Gandhinagar 382428, India }
\address {$^{93}$University of Szeged, D\'om t\'er 9, Szeged 6720, Hungary }
\address {$^{94}$Embry-Riddle Aeronautical University, Prescott, AZ 86301, USA }
\address {$^{95}$Tata Institute for Fundamental Research, Mumbai 400005, India }
\address {$^{96}$American University, Washington, D.C. 20016, USA }
\address {$^{97}$University of Massachusetts-Amherst, Amherst, MA 01003, USA }
\address {$^{98}$University of Adelaide, Adelaide, South Australia 5005, Australia }
\address {$^{99}$West Virginia University, Morgantown, WV 26506, USA }
\address {$^{100}$University of Bia{\l }ystok, 15-424 Bia{\l }ystok, Poland }
\address {$^{101}$SUPA, University of Strathclyde, Glasgow G1 1XQ, United Kingdom }
\address {$^{102}$IISER-TVM, CET Campus, Trivandrum Kerala 695016, India }
\address {$^{103}$Northwestern University, Evanston, IL 60208, USA }
\address {$^{104}$Institute of Applied Physics, Nizhny Novgorod, 603950, Russia }
\address {$^{105}$Pusan National University, Busan 609-735, Korea }
\address {$^{106}$Hanyang University, Seoul 133-791, Korea }
\address {$^{107}$NCBJ, 05-400 \'Swierk-Otwock, Poland }
\address {$^{108}$IM-PAN, 00-956 Warsaw, Poland }
\address {$^{109}$Rochester Institute of Technology, Rochester, NY 14623, USA }
\address {$^{110}$Monash University, Victoria 3800, Australia }
\address {$^{111}$Seoul National University, Seoul 151-742, Korea }
\address {$^{112}$ESPCI, CNRS, F-75005 Paris, France }
\address {$^{113}$Universit\`a di Camerino, Dipartimento di Fisica, I-62032 Camerino, Italy }
\address {$^{114}$Southern University and A\&M College, Baton Rouge, LA 70813, USA }
\address {$^{115}$College of William and Mary, Williamsburg, VA 23187, USA }
\address {$^{116}$Instituto de F\'\i sica Te\'orica, University Estadual Paulista/ICTP South American Institute for Fundamental Research, S\~ao Paulo SP 01140-070, Brazil }
\address {$^{117}$IISER-Kolkata, Mohanpur, West Bengal 741252, India }
\address {$^{118}$Rutherford Appleton Laboratory, HSIC, Chilton, Didcot, Oxon OX11 0QX, United Kingdom }
\address {$^{119}$Whitman College, 280 Boyer Ave, Walla Walla, WA 9936, USA }
\address {$^{120}$National Institute for Mathematical Sciences, Daejeon 305-390, Korea }
\address {$^{121}$Hobart and William Smith Colleges, Geneva, NY 14456, USA }
\address {$^{122}$Institute of Astronomy, 65-265 Zielona G\'ora, Poland }
\address {$^{123}$Andrews University, Berrien Springs, MI 49104, USA }
\address {$^{124}$Universit\`a di Siena, I-53100 Siena, Italy }
\address {$^{125}$Trinity University, San Antonio, TX 78212, USA }
\address {$^{126}$University of Washington, Seattle, WA 98195, USA }
\address {$^{127}$Abilene Christian University, Abilene, TX 79699, USA }



\date{12th February 2016}

\begin{abstract}
We present the results of a search for long-duration gravitational wave transients in two sets of data collected by the LIGO Hanford and LIGO Livingston detectors between November 5, 2005 and September 30, 2007, and July 7, 2009 and October 20, 2010, with a total observational time of \SfiveLivetime\, and \SsixLivetime, respectively.
The search targets gravitational wave transients of duration \unit[10--500]{s} in a frequency band of \unit[40--1000]{Hz}, with minimal assumptions about the signal waveform, polarization, source direction, or time of occurrence.
All candidate triggers were consistent with the expected background; as a result we set 90\% confidence upper limits on the rate of long-duration gravitational wave transients for different types of gravitational wave signals.
For signals from black hole accretion disk instabilities, we set upper limits on the source rate density between \unit[$3.4 \times 10^{-5}$--$9.4 \times 10^{-4}$]{Mpc$^{-3}$yr$^{-1}$} at 90\% confidence.
These are the first results from an all-sky search for unmodeled long-duration transient gravitational waves.
\end{abstract}


\maketitle


\section{\label{introduction}Introduction}

The goal of the Laser Interferometer Gravitational-Wave Observatory (LIGO)~\cite{Abbott:2007kv} and the Virgo detectors~\cite{Accadia:2012zzb} is to directly detect and study gravitational waves (GWs).
The direct detection of GWs holds the promise of testing general relativity in the strong-field regime, of providing a new probe of objects such as black holes and neutron stars, and of uncovering unanticipated new astrophysics.

LIGO and Virgo have jointly acquired data that have been used to search for many types of GW signals: unmodeled bursts of short duration ($<1$ s)~\cite{Abadie:2010mt,Abadie:2012rq,Aasi:2013vna,Abadie:2012prd,Aasi:2014iwa}, well-modeled chirps emitted by binary systems of compact objects~\cite{Abadie:2010yb,Abadie:2011nz,Abadie:2011kd,Aasi:2012rja,Aasi:2014bqj}, continuous signals emitted by asymmetric neutron stars~\cite{Abbott:2008fx,Aasi:2012fwa,Abadie:2011wj,Aasi:2013sia,Aasi:2013jya,Aasi:2014erp,Aasi:2013lva,Aasi:2014mtf}, as well as a stochastic background of GWs~\cite{Abbott:2011rs,Abbott:2011rr,Aasi:2014zwg,Aasi:2014jkh}.
For a complete review, see \cite{Bizouard:2013nxa}.
While no GW sources have been observed by the first-generation network of detectors, first detections are expected with the next generation of ground-based detectors: advanced LIGO~\cite{Aasi:2014jea}, advanced Virgo~\cite{Acernese:2014hva}, and the cryogenic detector KAGRA~\cite{Uchiyama:2014uza}.
It is expected that the advanced detectors, operating at design sensitivity, will be capable of detecting approximately 40 neutron star binary coalescences per year, although significant uncertainties exist~\cite{Aasi:2013wya}.

Previous searches for unmodeled bursts of GWs~\cite{Abadie:2010mt,Abadie:2012rq,Aasi:2013vna} targeted source objects such as core-collapse supernovae~\cite{Ott:2008wt}, neutron star to black hole collapse~\cite{Baiotti:2007np}, cosmic string cusps~\cite{Damour:2001bk}, binary black hole mergers~\cite{Belczynski:2014iua,lrr-2011-6,Etienne:2008xt}, star-quakes in magnetars~\cite{Mereghetti:2008je}, pulsar glitches~\cite{Andersson:2001xt}, and signals associated with gamma ray bursts (GRBs)~\cite{Briggs:2012ce}.
These burst searches typically look for signals of duration \unit[1]{s} or shorter.

At the other end of the spectrum, searches for persistent, unmodeled (stochastic) GW backgrounds have also been conducted, including isotropic~\cite{Abbott:2011rs}, anisotropic and point-source backgrounds~\cite{Abbott:2011rr}.
This leaves the parameter space of unmodeled transient GWs not fully explored; indeed, multiple proposed astrophysical scenarios predict long-duration GW transients lasting from a few seconds to hundreds of seconds, or even longer, as described in Section \ref{sec:sources}.
The first search for unmodeled long-duration GW transients was conducted using LIGO data from the S5 science run, in association with long GRBs~\cite{Aasi:2013cya}.
In this paper, we apply a similar technique~\cite{Thrane:2010ri} in order to search for long-lasting transient GW signals over all sky directions and for all times.
We utilize LIGO data from the LIGO Hanford and Livingston detectors from the S5 and S6 science runs, lasting from November 5, 2005 to September 30, 2007 and July 7, 2009 to October 20, 2010, respectively.

The organization of the paper is as follows.
In Section \ref{sec:sources}, we summarize different types of long-duration transient signals which may be observable by LIGO and Virgo.
In Section \ref{sec:dataset}, we describe the selection of the LIGO S5 and S6 science run data that have been used for this study.
We discuss the search algorithm, background estimation, and data quality methods in Section \ref{sec:search}.
In Section \ref{sec:sensitivity}, we evaluate the sensitivity of the search to simulated GW waveforms.
The results of the search are presented in Section \ref{sec:results}.
We conclude with possible improvements for a long-transient GW search using data from the advanced LIGO and Virgo detectors in Section \ref{sec:conclusion}.

\section{Astrophysical sources of long GW transients}\label{sec:sources}

Some of the most compelling astrophysical sources of long GW transients are associated with extremely complex dynamics and hydrodynamic instabilities following the collapse of a massive star's core in the context of core-collapse supernovae and long GRBs~\cite{Kotake:2011yv,Ott:2008wt,Thrane:2010ri}.
Soon after core collapse and the formation of a proto-neutron star, convective and other fluid instabilities (including standing accretion shock instability~\cite{Blondin:2002sm}) may develop behind the supernova shock wave as it transitions into an accretion shock.
In progenitor stars with rapidly rotating cores, long-lasting, non-axisymmetric rotational instabilities can be triggered by corotation modes~\cite{Ott:2005gj,ott:07prl,scheidegger:10b,Kuroda:2013rga}.
Long-duration GW signals are expected from these violently aspherical dynamics, following within tens of milliseconds of the short-duration GW burst signal from core bounce and proto-neutron star formation.
Given the turbulent and chaotic nature of post-bounce fluid dynamics, one expects a stochastic GW signal that could last from a fraction of a second to multiple seconds, and possibly even longer~\cite{Mueller:2003fs,Kotake:2009em,Ott:2008wt,mueller:13gw,yakunin:15,Thrane:2010ri}.

After the launch of an at least initially successful explosion, fallback accretion onto the newborn neutron star may spin it up, leading to non-axisymmetric deformation and a characteristic upward chirp signal (\unit[700]{Hz}--\unit[few]{kHz}) as the spin frequency of the neutron star increases over tens to hundreds of seconds~\cite{piro:11,Piro:2012ax}.
GW emission may eventually terminate when the neutron star collapses to a black hole.
The collapse process and formation of the black hole itself will also produce a short-duration GW burst~\cite{ott:11a,Baiotti:2007np,Dietrich:2014wja}.

In the collapsar model for long GRBs \cite{woosley:93}, a stellar-mass black hole forms, surrounded by a massive, self-gravitating accretion disk.
This disk may be susceptible to various non-axisymmetric hydrodynamic and magneto-hydrodynamic instabilities which may lead to fragmentation and inspiral of fragments into the central black hole (e.g.,~\cite{Piro:2006ja,Kiuchi:2011re}).
In an extreme scenario of such accretion disk instabilities (ADIs), magnetically ``suspended accretion'' is thought to extract spin energy from the black hole and dissipate it via GW emission from non-axisymmetric disk modes and fragments~\cite{vanPutten:2001sw,vanPutten:2003hd}.
The associated GW signal is potentially long-lasting (\unit[10--100]{s}) and predicted to exhibit a characteristic downward chirp.

Finally, in magnetar models for long and short GRBs (e.g.,~\cite{Metzger:2011aa,rowlinson:13}), a long-lasting post-GRB GW transient may be emitted by a magnetar undergoing rotational or magnetic non-axisymmetric deformation (e.g.,~\cite{corsi:09,gualteri:11}).

\section{Data selection}\label{sec:dataset}
During the fifth LIGO science run (S5, November 5, 2005 to September 30, 2007), the \unit[4]{km} and \unit[2]{km} detectors at Hanford, Washington (H1 and H2), and the \unit[4]{km} detector at Livingston, Louisiana (L1), recorded data for nearly two years.
They were joined on May 18, 2007 by the Virgo detector (V1) in Pisa, Italy, which was beginning its first science run.
After a two-year period of upgrades to the detectors and the decommissioning of H2, the sixth LIGO and second and third Virgo scientific runs were organized jointly from July 7, 2009 to October 10, 2010.

Among the four detectors, H1 and L1 achieved the best strain sensitivity, reaching $\approx 2 \times 10^{-23} / \sqrt{\text{Hz}}$ around \unit[150]{Hz} in 2010~\cite{LIGO:2012aa,Abadie:2010cg}.
Because of its reduced arm length, H2 sensitivity was at least a factor of 2 lower than H1 on average.
V1 sensitivity varied over time, but was always lower than the sensitivity of H1 and L1 by a factor between 1.5 and 5 at frequencies higher than \unit[60]{Hz}.
Moreover, the H1-L1 pair livetime was at least a factor 2 longer than the livetime of the H1-V1 and L1-V1 pairs added together.
Using Virgo data, however, could help with sky localization of source candidates; unfortunately, the sky localization was not implemented at the time of this search.
Consequently, including Virgo data in this analysis would have increased the overall search sensitivity by only a few percent or less at the cost of analyzing two additional pairs of detectors.
As a result, we have analyzed only S5 and S6 data from the H1-L1 pair for this search.

In terms of frequency content, we restrict the analysis to the \unit[40--1000]{Hz} band.
The lower limit is constrained by seismic noise, which steeply increases at lower frequencies in LIGO data.
The upper limit is set to include the most likely regions of frequency space for long-transient GWs, while keeping the computational requirements of the search at an acceptable level.
We note that the frequency range of our analysis includes the most sensitive band of the LIGO detectors, namely \unit[100--200]{Hz}.

Occasionally, the detectors are affected by instrumental problems (data acquisition failures, misalignment of optical cavities, etc.) or environmental conditions (bad weather, seismic motion, etc.) that decrease their sensitivity and increase the rate of data artifacts, or glitches.
Most of these periods have been identified and can be discarded from the analysis using data quality flags~\cite{Christensen:2010zz,Slutsky:2010ff,Aasi:2014mqd,Aasi:2012wd}. These are classified by each search into different categories depending on how the GW search is affected.

Category 1 data quality flags are used to define periods when the data should not be searched for GW signals because of serious problems, like invalid calibration.
To search for GW signals, the interferometers should be locked and there should be no evidence of environmental noise transients corrupting the measured signal.
For this search, we have used the category 1 data quality flags used by searches for an isotropic stochastic background of GWs~\cite{Abbott:2011rs,Aasi:2014zwg}.
This list of flags is almost identical to what has been used by the unmodeled all-sky searches for short-duration GW transients~\cite{Abadie:2010mt,Abadie:2012rq}.
We also discard times when simulated signals are injected into the detectors through the application of a differential force onto the mirrors.

Category 2 data quality flags are used to discard triggers which pass all selection cuts in a search, but are clearly associated with a detector malfunction or an environmental condition~\cite{Aasi:2014mqd}.
In Section \ref{subsec:DQ}, we explain which category 2 flags have been selected and how we use them in this search. 

Overall, we discard 5.8\% and 2.2\% of H1-L1 coincident data with our choices of category 1 data quality flags for S5 and S6, respectively.
The remaining coincident strain time series are divided into \unit[500]{s} intervals with 50\% overlap.
Intervals smaller than \unit[500]{s} are not considered.
For the H1-L1 pair, this results in a total observation time of \SfiveLivetime\,during S5 and \SsixLivetime\,for S6.

\section{Long transient GW search pipeline}\label{sec:search}

\subsection{Search algorithm}\label{subsec:algorithm}

The search algorithm we employ is based on the cross-correlation of data from two GW detectors, as described in \cite{Thrane:2010ri}.
This algorithm builds a frequency-time map ($ft$-map) of the cross-power computed from the strain time-series of two spatially separated detectors.
A pattern recognition algorithm is then used to identify clusters of above-threshold pixels in the map, thereby defining candidate triggers.
A similar algorithm has been used to search for long-lasting GW signals in coincidence with long GRBs in LIGO data~\cite{Aasi:2013cya}.
Here we extend the method to carry out an un-triggered (all-sky, all-time) search, considerably increasing the parameter space covered by previous searches.

Following~\cite{Thrane:2010ri}, each \unit[500]{s} interval of coincident data is divided into 50\% overlapping, Hann-windowed, \unit[1]{s} long segments.
Strain data from each detector in the given \unit[1]{s} segment are then Fourier transformed, allowing formation of $ft$-maps with a pixel size of \unit[1]{s} $\times$ \unit[1]{Hz}.
An estimator for GW power can be formed~\cite{Thrane:2010ri}:

\begin{equation}\label{eq:stamp_Y}
  \hat{Y}(t;f;\hat{\Omega}) = \frac{2}{\mathcal{N}}\text{Re}\left[ Q_{IJ}(t;f;\hat{\Omega})\tilde{s}_I^\star(t;f)\tilde{s}_J(t;f)\right].
\end{equation}
Here $t$ is the start time of the pixel, $f$ is the frequency of the pixel, $\hat{\Omega}$ is the sky direction, $\mathcal{N}$ is a window normalization factor, and $\tilde{s}_I$ and $\tilde{s}_J$ are the discrete Fourier transforms of the strain data from GW detectors $I$ and $J$. 
We use the LIGO H1 and L1 detectors as the $I$ and $J$ detectors, respectively.
The optimal filter $Q_{IJ}$ takes into account the phase delay due to the spatial separation of the two detectors, $\Delta\vec{x}_{IJ}$, and the direction-dependent efficiency of the detector pair, $\epsilon_{IJ}(t;\hat{\Omega})$:
\begin{equation}
Q_{IJ}(t;f;\hat{\Omega}) = \frac{e^{2\pi i f \Delta\vec{x}_{IJ} \cdot \hat{\Omega}/c}}{\epsilon_{IJ}(t;\hat{\Omega})}.
\end{equation}
The pair efficiency is defined by: 
\begin{equation}
\epsilon_{IJ}(t;\hat{\Omega})= \frac{1}{2}\nicesum_{A}{F_I^{A}(t;\hat{\Omega})F_J^{A}(t;\hat{\Omega})},
\end{equation}
where $F_I^{A}(t;\hat{\Omega})$ is the antenna factor for detector $I$ and $A$ is the polarization state of the incoming GW~\cite{Thrane:2010ri}.
An estimator for the variance of the $\hat{Y}(t;f,\hat{\Omega})$ statistic is then given by:
\begin{equation}\label{eq:sigma}
\hat{\sigma}_Y^2(t;f;\hat{\Omega}) = \frac{1}{2} \; |Q_{IJ}(t;f;\hat{\Omega})|^2 \; P_I^\text{adj}(t;f) \; P_J^\text{adj}(t;f),
\end{equation}
where $P_I^\text{adj}(t;f)$ is the average one-sided power spectrum for detector $I$, calculated by using the data in 8 non-overlapping segments on each side of time segment $t$~\cite{Thrane:2010ri}.
We can then define the cross-correlation signal-to-noise ratio (SNR) in a single pixel, $\rho$:
\begin{equation}\label{eq:rho}
  \rho(t;f;\hat{\Omega}) = \hat{Y}(t;f;\hat{\Omega}) / \hat{\sigma}_Y (t;f;\hat{\Omega}).
\end{equation}
Because this is proportional to strain squared, it is an energy SNR, rather than an amplitude SNR.





\begin{figure}[hbtp!]
  \centering
  \includegraphics[height=2.5in]{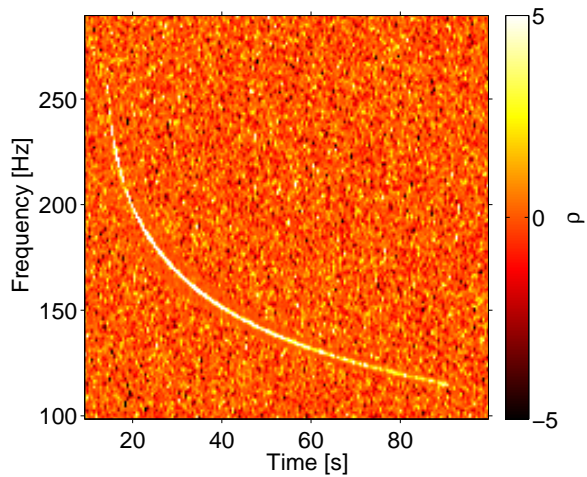}
  \caption{$ft$-map of $\rho$ (cross-correlation signal-to-noise ratio) using simulated Gaussian data.
    A simulated GW signal from an accretion disk instability~\cite{vanPutten:2001sw,vanPutten:2003hd} (model waveform ADI-E, see Table~\ref{tab:waveforms_ADI}) with known sky position is added to the data stream and is visible as a bright, narrow-band track. Blurring around the track is due to the usage of adjacent time segments in estimating $\hat{\sigma}_Y$; the estimate of $\hat{\sigma}_Y$ in these bins is affected by the presence of the GW signal.}
  \label{fig:snr_example}
\end{figure}

This statistic is designed such that true GW signals should induce positive definite $\rho$ when the correct filter is used (i.e., the sky direction $\hat{\Omega}$ is known).
Consequently, using a wrong sky direction in the filter results in reduced or even negative $\rho$ for real signals.
Figure~\ref{fig:snr_example} shows an example $ft$-map of $\rho$ containing a simulated GW signal with a known sky position.

Next, a seed-based clustering algorithm~\cite{burstegard} is applied to the $\rho$ $ft$-map to identify significant clusters of pixels.
In particular, the clustering algorithm applies a threshold of $|\rho| \geq 1$ to identify seed pixels, and then groups these seed pixels that are located within a fixed distance (two pixels) of each other into a cluster.
These parameters were determined through empirical testing with simulated long-transient GW signals similar to those used in this search (discussed further in Section~\ref{subsec:models}).
The resulting clusters (denoted $\Gamma$) are ranked using a weighted sum of the individual pixel values of $\hat{Y}$ and $\hat{\sigma}_Y$:
\begin{equation}\label{eq:snr_gamma}
  \text{SNR}_\Gamma(\hat{\Omega}) = \frac{\nicesum_{t;f\in\Gamma} \hat{Y}(t;f;\hat{\Omega})\hat{\sigma}_Y^{-2}(t;f;\hat{\Omega})}{\left(\nicesum_{t;f\in\Gamma} \hat{\sigma}_Y^{-2}(t;f;\hat{\Omega}) \right)^{1/2}}.
\end{equation}
$\text{SNR}_\Gamma(\hat{\Omega})$ represents the signal-to-noise ratio of the cluster $\Gamma$.

In principle, this pattern recognition algorithm could be applied for every sky direction $\hat{\Omega}$, since each sky direction is associated with a different filter $Q_{IJ}(t;f;\hat{\Omega})$.
However, this procedure is prohibitively expensive from a computational standpoint.
We have therefore modified the seed-based clustering algorithm to cluster both pixels with positive $\rho$ and those with negative $\rho$ (arising when an incorrect sky direction is used in the filter).
Since the sky direction is not known in an all-sky search, this modification allows for the recovery of some of the power that would normally be lost due to a suboptimal choice of sky direction in the filter.

The algorithm is applied to each $ft$-map a certain number of times, each iteration corresponding to a different sky direction.
The sky directions are chosen randomly, but are fixed for each stretch of uninterrupted science data.
Different methods for choosing the sky directions were studied, including using only sky directions where the detector network had high sensitivity and choosing the set of sky directions to span the set of possible signal time delays.
The results indicated that sky direction choice did not have a significant impact on the sensitivity of the search.

We also studied the effect that the number of sky directions used had on the search sensitivity.
We found that the search sensitivity increased approximately logarithmically with the number of sky directions, while the computational time increased linearly with the number of sky directions.
The results of our empirical studies indicated that using five sky directions gave the optimal balance between computational time and search sensitivity.

This clustering strategy results in a loss of sensitivity of $\approx$10--20\% for the waveforms considered in this search as compared to a strategy using hundreds of sky directions and clustering only positive pixels.
However, this strategy increases the computational speed of the search by a factor of 100 and is necessary to make the search computationally feasible.

We also apply two data cleaning techniques concurrently with the data processing.
First, we remove frequency bins that are known to be contaminated by instrumental and environmental effects.
This includes the violin resonance modes of the suspensions, power line harmonics, and sinusoidal signals injected for calibration purposes.
In total, we removed 47 \unit[1]{Hz}-wide frequency bins from the S5 data, and 64 \unit[1]{Hz}-wide frequency bins from the S6 data.
Second, we require the waveforms observed by the two detectors to be consistent with each other, so as to suppress instrumental artifacts (glitches) that affect only one of the detectors.
This is achieved by the use of a consistency-check algorithm~\cite{stamp_glitch} which compares the power spectra from each detector, taking into account the antenna factors.

\subsection{Background estimation}\label{subsec:background}

An important aspect of any GW search is understanding the background of accidental triggers due to detector noise; this is crucial for preventing false identification of noise triggers as GW candidates.
To estimate the false alarm rate (FAR), i.e. the rate of accidental triggers due to detector noise, we introduce a non-physical time-shift between the H1 and L1 strain data before computing $\rho$.
Analysis of the time-shifted data proceeds identically to that of unshifted data (see Section~\ref{subsec:algorithm} for more details).
With this technique, and assuming the number of hypothetical GW signals is small, the data should not contain a correlated GW signal, so any triggers will be generated by the detector noise.
We repeat this process for multiple time-shifts in order to gain a more accurate estimate of the FAR from detector noise.

As described in Section \ref{sec:dataset}, each analysis segment is divided into \unit[500]{s} long intervals which overlap by 50\% and span the entire dataset.
For a given time-shift $i$, the H1 data from interval $n$ are correlated with L1 data from the interval $n+i$.
Since the time gap between two consecutive intervals may be non-zero, the actual time-shift applied in this process is at least $500 \times i$ seconds.
The time-shift is also circular: if for a time-shift $i$, $n+i>N$ (where $N$ is the number of overlapping intervals required to span the dataset), then H1 data from the interval $n$ are correlated with L1 data from the interval $n+i-N$.
It is important to note that the minimum time-shift duration is much longer than the light travel time between the two detectors and also longer than the signal models we consider (see Section~\ref{sec:sensitivity} for more information) in order to prevent accidental correlations.

Using this method, 100 time-shifts have been processed to estimate the background during S5, amounting to a total analyzed livetime of \SfiveTSLivetime.
We have also studied 100 time-shifts of S6 data, with a total analyzed livetime of \SsixTSLivetime.
The cumulative rates of background triggers for the S5 and S6 datasets can be seen in Figures~\ref{fig:FAR_S5} and \ref{fig:FAR_S6}, respectively.

\subsection{Rejection of loud noise triggers}\label{subsec:DQ}

\begin{figure}[hbtp!]
  \centering
  \includegraphics[height=2.5in]{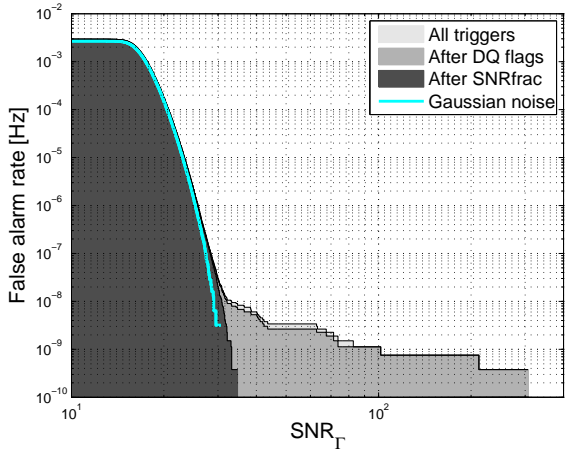}
  \caption{The false alarm rate is shown as a function of the trigger $\text{SNR}_{\Gamma}$, for 100 time-shifts of data from the S5 science run. Distributions are shown before and after applying the post-processing cuts. Here, SNRfrac refers to a post-processing cut based on how a trigger's power is distributed in time (described further in Section~\ref{subsec:DQ}). Also shown is the FAR distribution generated by a Monte Carlo simulation assuming Gaussian detector noise. Recall that this is an energy SNR, rather than an amplitude SNR.}
  \label{fig:FAR_S5}
\end{figure}

\begin{figure}[hbtp!]
  \centering
  \includegraphics[height=2.5in]{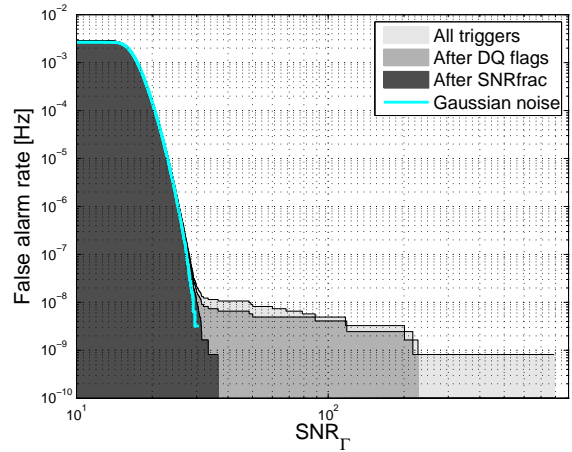}
  \caption{The false alarm rate is shown as a function of the trigger signal-to-noise ratio, $\text{SNR}_{\Gamma}$, for 100 time-shifts of data from the S6 science run. See caption of Figure~\ref{fig:FAR_S5} for all details.}
  \label{fig:FAR_S6}
\end{figure}

As shown in Figures~\ref{fig:FAR_S5} and \ref{fig:FAR_S6}, the background FAR distribution has a long tail extending to $\text{SNR}_{\Gamma} > 100$; this implies that detector noise alone can generate triggers containing significant power.
Many of these triggers are caused by short bursts of non-stationary noise (glitches) in H1 and/or L1, which randomly coincide during the time-shifting procedure.
It is important to suppress these types of triggers so as to improve the significance of true GW signals in the unshifted data.

These glitches are typically much less than \unit[1]{s} in duration, and as a result, nearly all of their power is concentrated in a single \unit[1]{s} segment.
To suppress these glitches, we have defined a discriminant variable, SNRfrac, that measures the fraction of $\text{SNR}_\Gamma$ located in a single time segment.
The same SNRfrac threshold of 0.45 was found to be optimal for all simulated GW waveforms using both S5 and S6 data.
This threshold was determined by maximizing the search sensitivity for a set of simulated GW signals (see Section~\ref{sec:sensitivity}); we note that this was done before examining the unshifted data, using only time-shifted triggers and simulated GW signals. The detection efficiency is minimally affected (less than 1\%) by this SNRfrac threshold choice. 


We also utilize LIGO data quality flags to veto triggers generated by a clearly identified source of noise.
We have considered all category 2 data quality flags used in unmodeled or modeled transient GW searches~\cite{Aasi:2014mqd}.
These flags were defined using a variety of environmental monitors (microphones, seismometers, magnetometers, etc.) and interferometer control signals to identify stretches of data which may be compromised due to local environmental effects or instrumental malfunction.
Since many of these data quality flags are not useful for rejecting noise triggers in this analysis, we select a set of effective data quality flags by estimating the statistical significance of the coincidence between these data quality flags and the 100 loudest triggers from the time-shifted background study (no SNRfrac selection applied). The significance is defined by comparing the number of coincident triggers with the accidental coincidence mean and standard deviation.
Given the small number of triggers we are considering (100), and in order to avoid accidental coincidence, we have applied a stringent selection: only those data quality flags which have a statistical significance higher than 12 standard deviations (as defined above) and an efficiency over deadtime ratio larger than 8 have been selected. 
Here, efficiency refers to the fraction of noise triggers flagged, while deadtime is the amount of science data excluded by the flag.

For both the S5 and S6 datasets, this procedure selected data quality flags which relate to malfunctions of the longitudinal control of the Fabry-Perot cavities and those which indicate an increase in seismic noise.
The total deadtime which results from applying these data quality (DQ) flags amounts to $\approx$12 hours in H1 and L1 ($0.18\%$) for S5 and $\approx$4 hours in H1 ($0.13\%$) and $\approx$7 hours in L1 ($0.22\%$) for S6.


As shown in Figures~\ref{fig:FAR_S5} and \ref{fig:FAR_S6}, these two data quality cuts (SNRfrac and DQ flags) are useful for suppressing the high-SNR$_\Gamma$ tail of the FAR distribution.
More precisely, the SNRfrac cut is very effective for cleaning up coincident glitches with high $\text{SNR}_\Gamma$, while the DQ flags are capable of removing less extreme triggers occurring due to the presence of a well-identified noise source.
We have thus decided to look at the unshifted (zero-lag) triggers after the SNRfrac cut is applied and reserve the DQ flags for the exclusion of potential GW candidates that are actually due to a well-understood instrumental problem. 

After the application of the SNRfrac cut, the resulting FAR distribution can be compared with that of a Monte Carlo simulation using a Gaussian noise distribution (assuming an initial LIGO noise sensitivity curve).
A discrepancy of $\approx$10\% in the total number of triggers and a slight excess of loud triggers are observed for both the S5 and S6 datasets when compared to Gaussian noise.

\section{Search sensitivity}\label{sec:sensitivity}

\subsection{GW signal models}\label{subsec:models}
To assess the sensitivity of our search to realistic GW signals, we use 15 types of simulated GW signals.
Four of these waveforms are based on an astrophysical model of a black hole accretion disk instability~\cite{vanPutten:2001sw,vanPutten:2003hd}.
The other 11 waveforms are not based on a model of an astrophysical GW source, but are chosen to encapsulate various characteristics that long-transient GW signals may possess, including duration, frequency content, bandwidth, and rate of change of frequency.
These {\it ad hoc} waveforms can be divided into three families: sinusoids with a time-dependent frequency content, sine-Gaussians, and band-limited white noise bursts.
All waveforms have \unit[1]{s}-long Hann-like tapers applied to the beginning and end of the waveforms in order to prevent data artifacts which may occur when the simulated signals have high intensity.
In this section, we give a brief description of each type of GW signal model.

\subsubsection{Accretion disk instabilities}\label{subsubsec:adi}
In this study, we include four variations on the ADI model (see Section~\ref{sec:sources} for more details).
Although this set of waveforms does not span the entire parameter space of the ADI model, it does encapsulate most of the possible variations in the signal morphology in terms of signal durations, frequency ranges and derivatives, and amplitudes (see Table~\ref{tab:waveforms_ADI} for a summary of the waveforms).
While these waveforms may not be precise representations of realistic signals, they capture the salient features of many proposed models and produce long-lived spectrogram tracks.

  \begin{center}
  \begin{table}[hbtp!]
    \begin{tabular}{c  c  c  c  c  c}
      \hline\hline
      Waveform & $M [M_{\odot}]$ & $a^*$ & $\epsilon$ & Duration [s] & Frequency [Hz] \\\hline
      ADI-A & 	 5	&  0.30	&  0.050 & 39	& 135--166	  \\\hline
      ADI-B & 	 10	&  0.95	&  0.200 & 9	& 110--209	  \\\hline
      ADI-C & 	 10	&  0.95	&  0.040 & 236	& 130--251	  \\\hline
      ADI-E & 	 8	&  0.99	&  0.065 & 76	& 111--234	  \\\hline
      \hline
    \end{tabular}
    \caption{List of ADI waveforms~\cite{vanPutten:2001sw,vanPutten:2003hd} used to test the sensitivity of the search. Here, $M$ is the mass of the central black hole, $a^*$ is the dimensionless Kerr spin parameter of the black hole, and $\epsilon$ is the fraction of the disk mass that forms clumps.  Frequency refers to the ending and starting frequencies of the GW signal, respectively.  All waveforms have an accretion disk mass of 1.5 $M_\odot$. }
    \label{tab:waveforms_ADI}
  \end{table}
  \end{center}

\subsubsection{Sinusoids}\label{subsubsec:sinusoid}

The sinusoidal waveforms are characterized by a sine function with a time-dependent frequency content.
The waveforms are described by

\begin{align}\label{eq:sinusoid_plus}
  h_+(t) = &\frac{1+\cos^2\iota}{2} \cos2\psi \cos\phi(t) - \nonumber\\&\cos\iota \sin2\psi \sin\phi(t),
\end{align}
\begin{align}\label{eq:sinusoid_cross}
  h_\times(t) = &\frac{1+\cos^2\iota}{2} \sin2\psi \cos\phi(t) + \nonumber\\&\cos\iota \cos2\psi \sin\phi(t),
\end{align}
where $\iota$ is the inclination angle of the source, $\psi$ is the source polarization, and $\phi(t)$ is a phase time-series, given by

\begin{equation}\label{eq:f_t}
  \phi(t) = 2\pi \left(f_0t + \frac{1}{2}\left(\frac{df}{dt}\right)t^2 + \frac{1}{6}\left(\frac{d^2 f}{dt^2}\right)t^3 \right).
\end{equation}

Two of the waveforms are completely monochromatic, two have a linear frequency dependence on time, and two have a quadratic frequency dependence on time.
This family of waveforms is summarized in Table~\ref{tab:waveforms_sine}.

\begin{center}
  \begin{table}[hbtp!]
    \begin{tabular}{c  c  c  c  c}
      \hline\hline
      Waveform & Duration [s] & $f_0$ [Hz] & $\frac{df}{dt}$ [Hz/s] & $\frac{d^2 f}{dt^2}$ [Hz/s$^2$] \\\hline
      MONO-A   & 150          & 90         & 0.0                    & 0.00                            \\\hline
      MONO-B   & 250          & 505        & 0.0                    & 0.00                            \\\hline
      LINE-A   & 250          & 50         & 0.6                    & 0.00                            \\\hline
      LINE-B   & 100          & 900        & -2.0                   & 0.00                            \\\hline
      QUAD-A   & 30           & 50         & 0.0                    & 0.33                            \\\hline
      QUAD-B   & 70           & 500        & 0.0                    & 0.04                            \\\hline
      \hline
    \end{tabular}
    \caption{List of sinusoidal waveforms used to test the sensitivity of the search. Here, $f_0$ is the initial frequency of the signal, $df/dt$ is the frequency derivative, and $d^2f/dt^2$ is the second derivative of the frequency.}
    \label{tab:waveforms_sine}
  \end{table}
\end{center}

\subsubsection{Sine-Gaussians}\label{subsubsec:sg}
The sine-Gaussian waveforms are essentially monochromatic signals (see Equations~\ref{eq:sinusoid_plus} and \ref{eq:sinusoid_cross}) multiplied by a Gaussian envelope:
\begin{equation}\label{eq:sg}
  e^{-t^2/\tau^2}.
\end{equation}
Here, $\tau$ is the decay time, which defines the width of the Gaussian envelope.
This set of waveforms is summarized in Table~\ref{tab:waveforms_sg}.

\begin{center}
  \begin{table}[hbtp!]
    \begin{tabular}{c  c  c  c}
      \hline\hline
      Waveform & Duration [s] & $f_0$ [Hz]     & $\tau$ [s] \\\hline
      SG-A     & 150          & 90             & 30         \\\hline
      SG-B     & 250          & 505            & 50         \\\hline
      \hline
    \end{tabular}
    \caption{List of sine-Gaussian waveforms used to test the sensitivity of the search. Here, $\tau$ is the decay time of the Gaussian envelope.}
    \label{tab:waveforms_sg}
  \end{table}
\end{center}

\subsubsection{Band-limited white noise bursts}\label{subsubsec:wnb}

We have generated white noise and used a 6th order Butterworth band-pass filter to restrict the noise to the desired frequency band.
Each polarization component of the simulated waveforms is generated independently; thus, the two components are uncorrelated.
This family of waveforms is summarized in Table~\ref{tab:waveforms_wnb}.

\begin{center}
  \begin{table}[hbtp!]
    \begin{tabular}{c  c  c}
      \hline\hline
      Waveform & Duration [s] & Frequency band [Hz] \\\hline
      WNB-A    & 20           & 50--400           \\\hline
      WNB-B    & 60           & 300--350          \\\hline
      WNB-C    & 100          & 700--750          \\\hline
      \hline
    \end{tabular}
    \caption{List of band-limited white noise burst waveforms used to test the sensitivity of the search.}
    \label{tab:waveforms_wnb}
  \end{table}
\end{center}

\subsection{Sensitivity study}\label{subsec:sens_study}

Using the waveforms described in the previous section, we performed a sensitivity study to determine the overall detection efficiency of the search as a function of waveform amplitude.
First, for each of the 15 models, we generated 1500 injection times randomly between the beginning and the end of each of the two datasets, such that the injected waveform was fully included in a group of at least one 500 second-long analysis window.
A minimal time lapse of \unit[1000]{s} between two injections was enforced. 
For each of the 1500 injection times, we generated a simulated signal with random sky position, source inclination, and waveform polarization angle. 
The time-shifted data plus simulated signal was then analyzed using the search algorithm described in Section~\ref{sec:search}.
The simulated signal was considered recovered if the search algorithm found a trigger fulfilling the following requirements:
\begin{itemize}
\itemsep2pt
\item The trigger was found between the known start and end time of the simulated signal.
\item The trigger was found within the known frequency band of the signal.
\item The $\text{SNR}_\Gamma$ of the trigger exceeded a threshold determined by the loudest trigger found in each dataset (using the unshifted data).
\end{itemize}
This was repeated with 16 different signal amplitudes (logarithmically spaced) for each waveform and injection time in order to fully characterize the search's detection efficiency as a function of signal strength.

\begin{figure*}[hbtp!]
  \centering
  \subfloat[]{\includegraphics[height=2.2in]{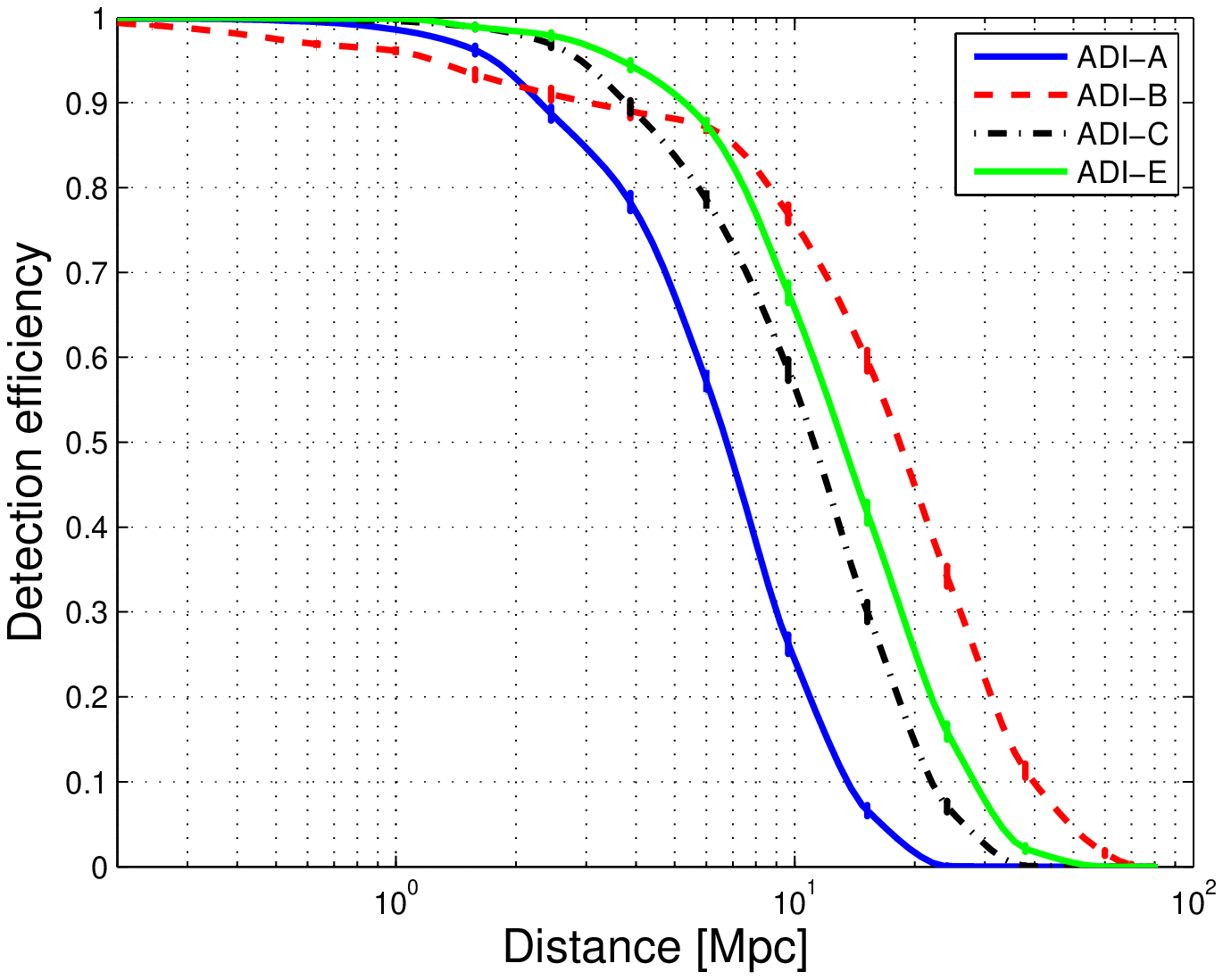}} \qquad
  \subfloat[]{\includegraphics[height=2.2in]{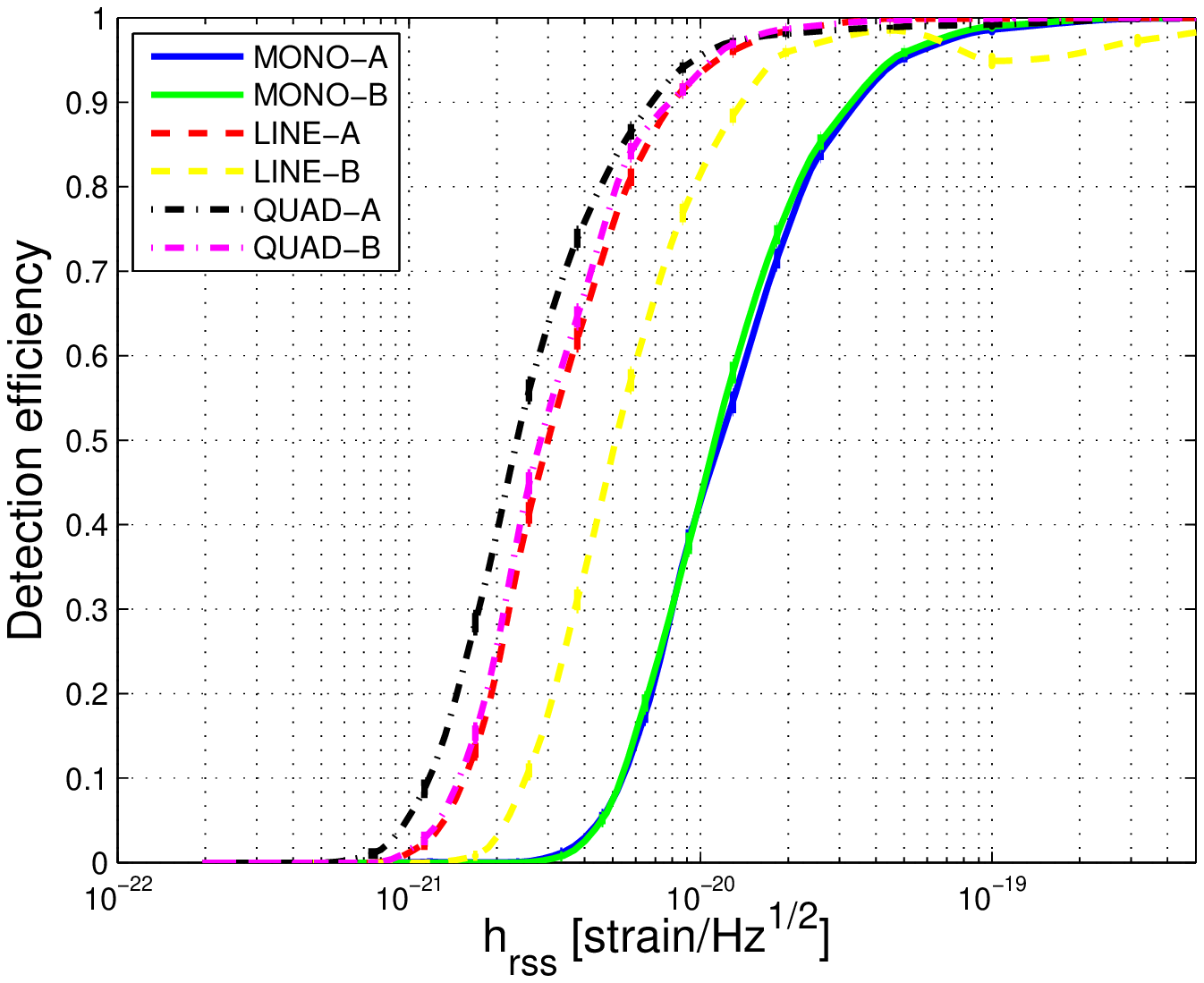}} \\
  \subfloat[]{\includegraphics[height=2.2in]{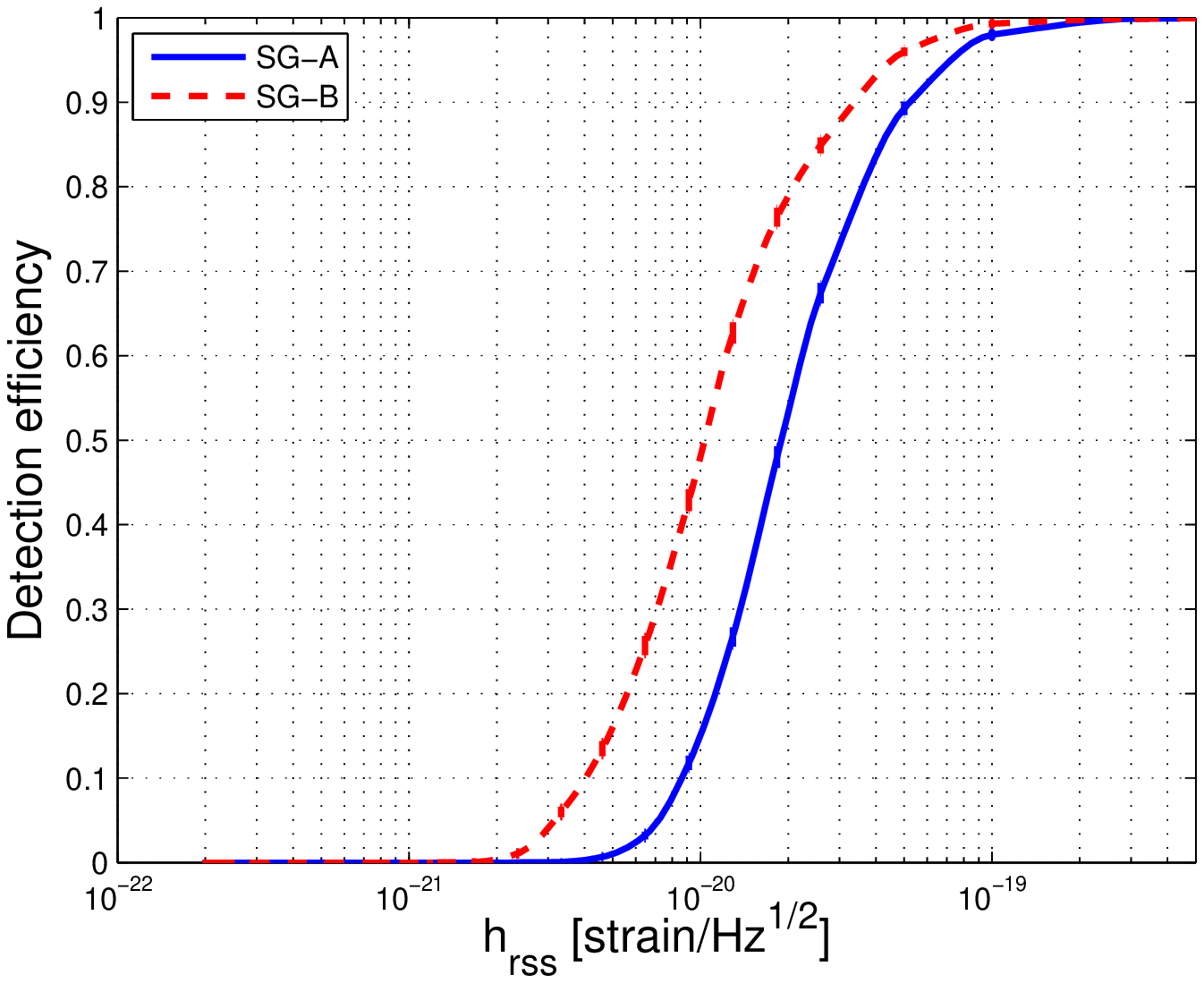}} \qquad
  \subfloat[]{\includegraphics[height=2.2in]{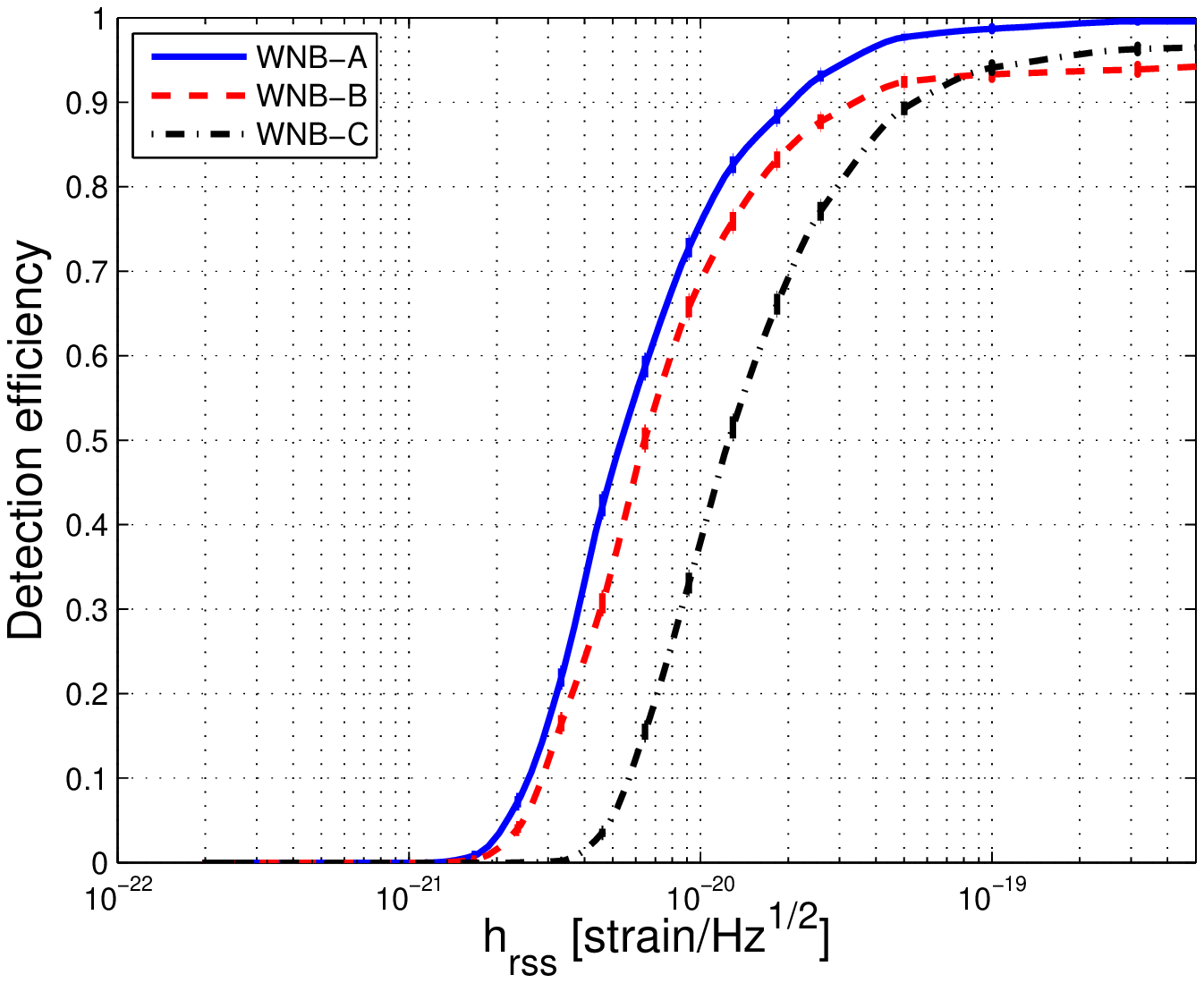}} \\
  \caption{Efficiency of the search pipeline at recovering different waveforms as a function of the distance to the source (for ADI waveforms) or the signal strength $h_\text{rss}$ (all others).  All results shown here used data from the S6 science run. The SNRfrac threshold is set at 0.45 and the recovery threshold is set at $\text{SNR}_\Gamma = 27.13$. The error bars are computed using binomial statistics.
    Top left: ADI waveforms.
    Top right: sinusoidal waveforms.
    Bottom left: sine-Gaussian waveforms.
    Bottom right: white noise burst waveforms.}
  \label{fig:efficiency}
\end{figure*}

In Figure~\ref{fig:efficiency}, we show the efficiency, or ratio of recovered signals to the total number of simulations, as a function of either the distance to the source or the root-sum-squared strain amplitude ($h_\text{rss}$) arriving at the Earth, defined as
\begin{equation}\label{eq:hrss}
h_\text{rss} = \sqrt{\int{\left( |h_{+} (t)|^2 + |h_{\times}(t)|^2 \right)dt}}.
\end{equation}

Among each family of waveforms, the pipeline efficiency has a frequency dependence that follows the data strain sensitivity of the detectors.
The duration of the signal also plays a role, but to a lesser extent.
We also note that the search efficiency for monochromatic waveforms (MONO and SG) is significantly worse than for the other waveforms.
This is due to the usage of adjacent time segments to compute $\hat{\sigma}_Y$ (see Equation~\ref{eq:sigma}), which is affected by the presence of the GW signal.

\section{Results}\label{sec:results}
\begin{figure}[hbtp!]
  \centering
  \includegraphics[height=2.5in]{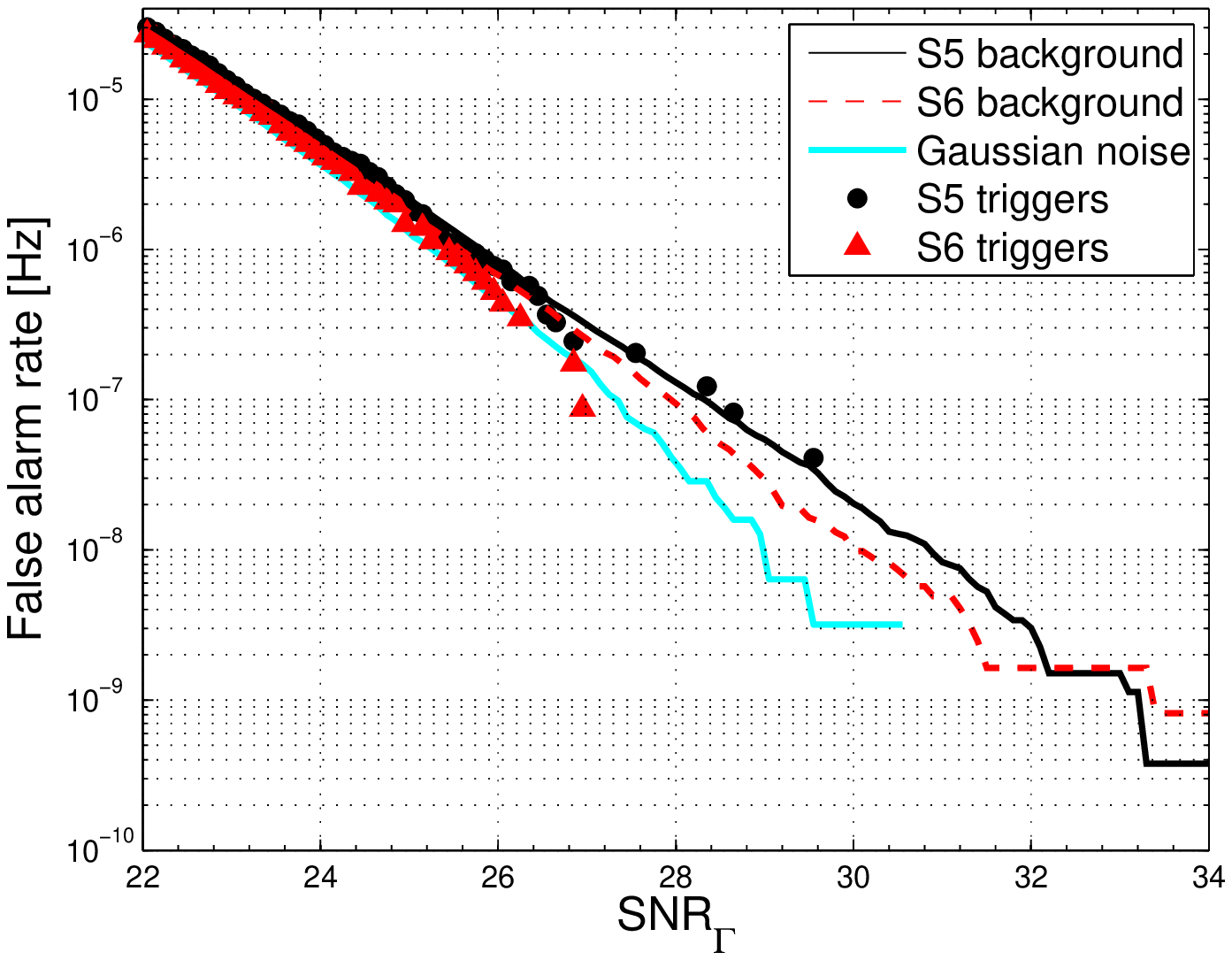}
  \caption{FAR distribution of unshifted triggers from S5 (black circles) and S6 (red triangles) as a function of the trigger signal-to-noise ratio, $\text{SNR}_{\Gamma}$. The distributions are compared to the estimated background distributions for the S5 (solid black) and S6 (dashed red) datasets. We observe a slight deficit of triggers in S6 that remains within one standard deviation of what is expected from the time-shifted triggers. Also shown is the FAR distribution generated by a Monte Carlo simulation assuming Gaussian detector noise (solid cyan).}
  \label{fig:FAR_S5_S6_ZL}
\end{figure}

Having studied the background triggers and optimized the SNRfrac threshold using both background and simulated signals, the final step in the analysis is to apply the search algorithm to the unshifted (zero-lag) data (i.e. zero time-shift between the H1 and L1 strain time series) in order to search for GW candidates. 
The resulting distributions of $\text{SNR}_\Gamma$ for the zero-lag S5 and S6 datasets are compared to the corresponding background trigger distributions in Figure~\ref{fig:FAR_S5_S6_ZL}.
A slight deficit of triggers is present in the S6 zero-lag, but it remains within one standard deviation of what is expected from the background.

\subsection{Loudest triggers}\label{subsec:candidates}
Here, we consider the most significant triggers from the S5 and S6 zero-lag analyses.
We check the FAR of each trigger, which is the number of background triggers with $\text{SNR}_\Gamma$ larger than a given threshold $\text{SNR}_\Gamma^\star$ divided by the total background livetime, $T_\text{bkg}$.
We also consider the false alarm probability (FAP), or the probability of observing at least $N$ background triggers with $\text{SNR}_\Gamma$ higher than $\text{SNR}_\Gamma^\star$:
\begin{equation}\label{eq:FAP}
  \text{FAP}(N) = 1 - \nicesum_{n=0}^{n=N-1} \frac{\mu_\text{bkg}^n}{n!} \times e^{-\mu_\text{bkg}},
\end{equation}
where $\mu_\text{bkg}$ is the number of background triggers expected from a Poisson process (given by $\mu_\text{bkg}=T_\text{obs} \times \text{FAR}(\text{SNR}_\Gamma^\star)$ where $T_{\text{obs}}$ is the observation time). 
For the loudest triggers in each dataset, we take $N=1$ to estimate the FAP.

The most significant triggers from the S5 and S6 zero-lag analyses occurred with false alarm probabilities of 54\% and 92\%, respectively.
They have respective false alarm rates of \unit[1.00]{yr$^{-1}$} and \unit[6.94]{yr$^{-1}$}.
This shows that triggers of this significance are frequently generated by detector noise alone, and thus, these triggers cannot be considered GW candidates.

Additional follow-up indicated that no category 2 data quality flags in H1 nor L1 were active at the time of these triggers.
The examination of the $ft$-maps, the whitened time series around the time of the triggers, and the monitoring records indicate that these triggers were due to a small excess of noise in H1 and/or L1, and are not associated with a well-identified source of noise.
More information about these triggers is provided in Table~\ref{tab:events}.

\begin{table}[hbtp!]
  \centering
  \begin{tabular}{ c c c c c c}
    \hline\hline
    Dataset & $\text{SNR}_\Gamma$ & FAR [yr$^{-1}$] & FAP & GPS time & Freq. [Hz] \\\hline
    S5 & 29.65 & 1.00 & 0.54 & 851136555.0 & 129--201 \\\hline
    S6 & 27.13 & 6.94 & 0.92 & 958158359.5 & 537--645 \\\hline
    \hline
  \end{tabular}
  \caption{The most significant triggers from the S5 and S6 datasets.  GPS times given correspond to trigger start times; both triggers had durations of 23.5 seconds.}
  \label{tab:events}
\end{table}

\subsection{Rate upper limits}\label{subsec:rates}
Since no GW candidates were identified, we proceed to place upper limits on the rate of long-duration GW transients assuming an isotropic and uniform distribution of sources.
We use two implementations of the loudest event statistic~\cite{Biswas:2009cqg} to set upper limits at 90\% confidence.
The first method uses the false alarm density (FAD) formalism~\cite{Abadie:2011kd, Abadie:2012prd}, which accounts for both the background detector noise and the sensitivity of the search to simulated GW signals.
For each simulated signal model, one calculates the efficiency of the search as a function of source distance at a given threshold on $\text{SNR}_\Gamma$, then integrates the efficiency over volume to gain a measure of the volume of space which is accessible to the search (referred to as the visible volume).
The threshold is given by the $\text{SNR}_\Gamma$ of the ``loudest event,'' or in this case, the trigger with the lowest FAD.
The 90\% confidence upper limits are calculated (following~\cite{Biswas:2009cqg}) as
\begin{equation}\label{eq:LES_vvis}
R_{90\%,\text{VT}} = \frac{2.3}{\nicesum_k V_{\text{vis},k} (\text{FAD}^\star) \times T_{\text{obs},k}}.
\end{equation}
Here, the index $k$ runs over datasets, $V_{\text{vis},k} (\text{FAD}^\star)$ is the visible volume of search $k$ calculated at the FAD of the loudest zero-lag trigger ($\text{FAD}^\star$), and $T_{\text{obs},k}$ is the observation time, or zero-lag livetime of search $k$.
The factor of 2.3 in the numerator is the mean rate (for zero observed triggers) which should give a non-zero number of triggers 90\% of the time; it can be calculated by solving Equation~\ref{eq:FAP} for $\mu_\text{bkg}$ with $N=1$ and a FAP of 0.9 (i.e., $1 - e^{-\mu_\text{bkg}} = 0.9$).
The subscript $\text{VT}$ indicates that these upper limits are in terms of number of observations per volume per time.

Due to the dependence of these rate upper limits on distance to the source, they cannot be calculated for the {\it ad hoc} waveforms without setting an arbitrary source distance.
A full description of the FAD and visible volume formalism is given in Appendix~\ref{sec:vvis_FAD}; in Table~\ref{tab:ADI_FAD_UL}, we present these upper limits for the four ADI waveforms.

\begin{table}[hbtp!]
  \centering
  \begin{tabular}{ c c c c }
    \hline\hline
    Waveform & \multicolumn{2}{ c }{$V_\text{vis}$ [Mpc$^3$]} & $R_{90\%,\text{VT}}$ [Mpc$^{-3}$yr$^{-1}$] \\
    & S5 & S6 & \\\hline
    ADI-A & $1.8 \times 10^3$ & $3.6 \times 10^3$ & $9.4 \times 10^{-4}$ \\\hline
    ADI-B & $5.7 \times 10^4$ & $9.1 \times 10^4$ & $3.4 \times 10^{-5}$ \\\hline
    ADI-C & $7.8 \times 10^3$ & $1.6 \times 10^4$ & $2.2 \times 10^{-4}$ \\\hline
    ADI-E & $1.6 \times 10^4$ & $3.2 \times 10^4$ & $1.1 \times 10^{-4}$ \\\hline
    \hline
  \end{tabular}
  \caption{Rate upper limits on ADI waveforms calculated with Equation~\ref{eq:LES_vvis}.  The visible volume of each search is shown to illustrate relative search sensitivities.  Uncertainties on the visible volumes are not included, but are primarily due to calibration uncertainty, and are 53\% and 24\% for all waveforms in S5 and S6, respectively.  1 $\sigma$ uncertainties on the upper limits are marginalized over using the Bayesian method described in Section~\ref{sec:bayesian_FAD}.}
  \label{tab:ADI_FAD_UL}
\end{table}

Statistical and systematic uncertainties are discussed further in Appendix~\ref{sec:vvis_FAD}, but we note here that the dominant source of uncertainty is the amplitude calibration uncertainty of the detectors.
During the S5 science run, the amplitude calibration uncertainty was measured to be 10.4\% and 14.4\% for the H1 and L1 detectors, respectively, in the \unit[40--2000]{Hz} frequency band~\cite{Abadie:2010px}.
Summing these uncertainties in quadrature gives a total calibration uncertainty of 17.8\% on the amplitude, and thus, an uncertainty of 53.4\% on the visible volume.
For S6, the amplitude calibration uncertainty was measured at 4.0\% for H1 and 7.0\% for L1 in the \unit[40--2000]{Hz} band, resulting in a total calibration uncertainty of 8.1\% on the amplitude and 24.2\% on the visible volume~\cite{LIGO:S6calib}.
These uncertainties are marginalized over using a Bayesian method discussed in Appendix~\ref{sec:bayesian_FAD}.
The upper limits presented in Table~\ref{tab:ADI_FAD_UL} are conservative and include these uncertainties.

\begin{figure*}[hbtp!]
  \centering
  \subfloat[]{\includegraphics[height=2.2in]{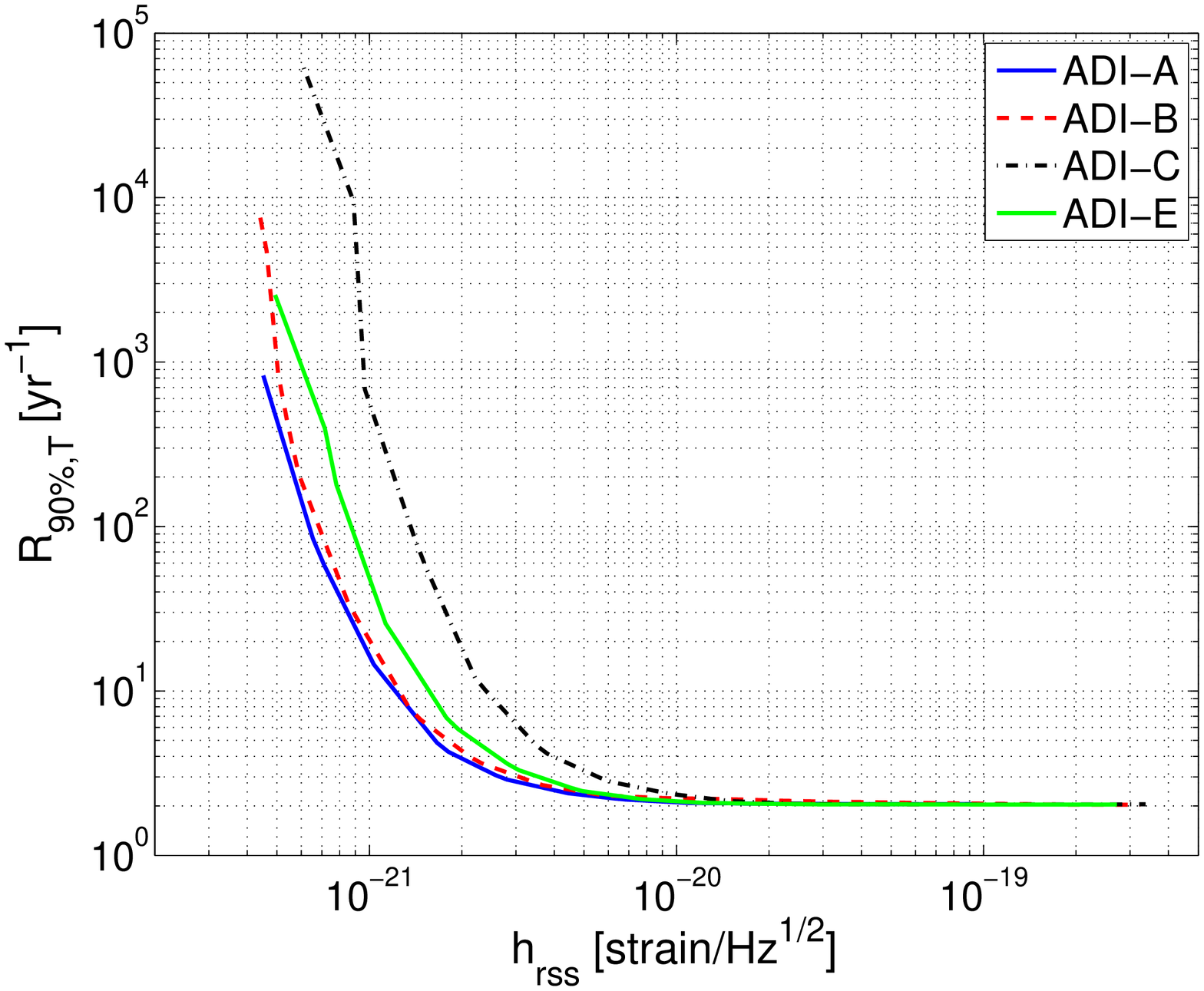}} \qquad\qquad
  \subfloat[]{\includegraphics[height=2.2in]{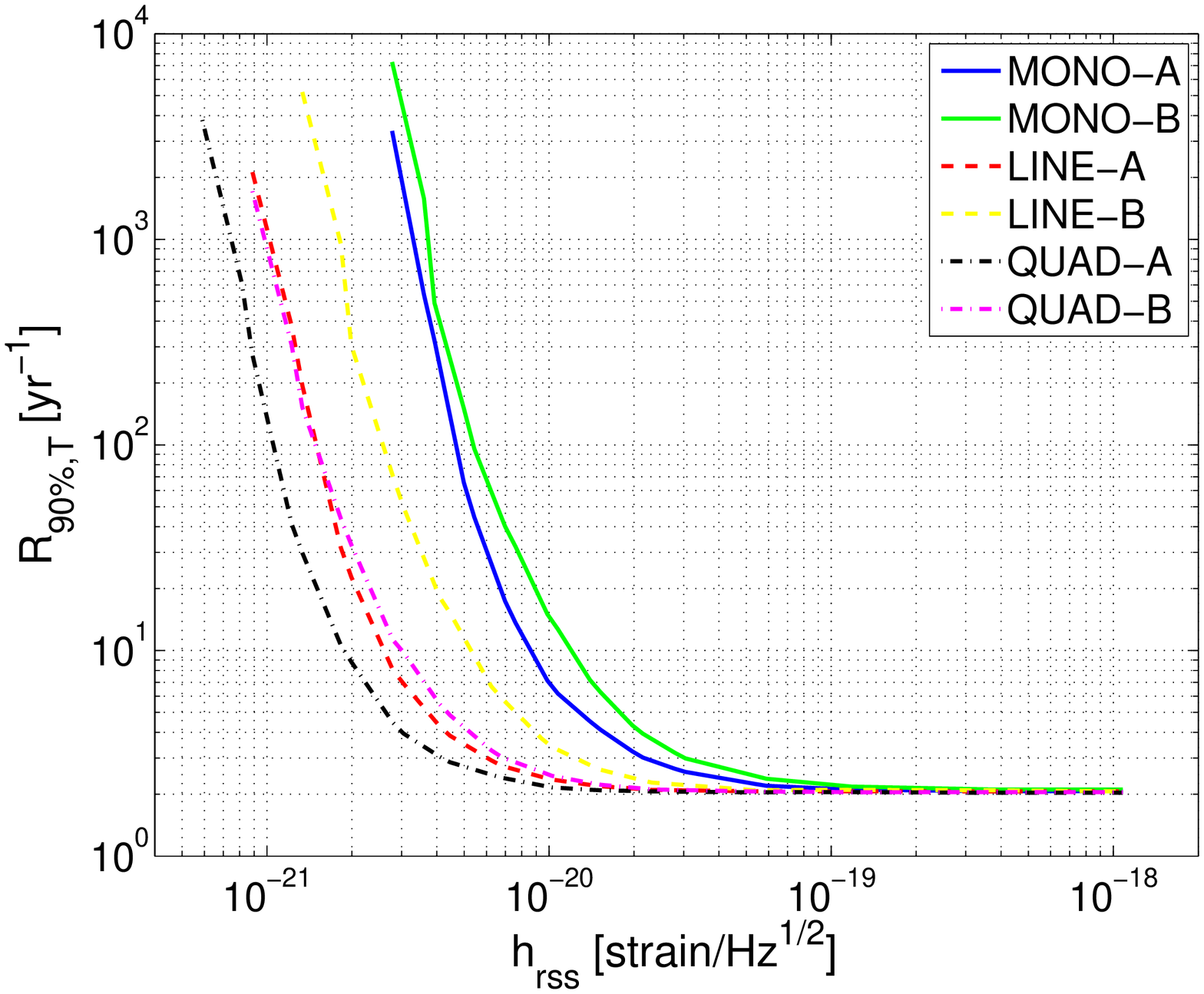}} \\
  \subfloat[]{\includegraphics[height=2.2in]{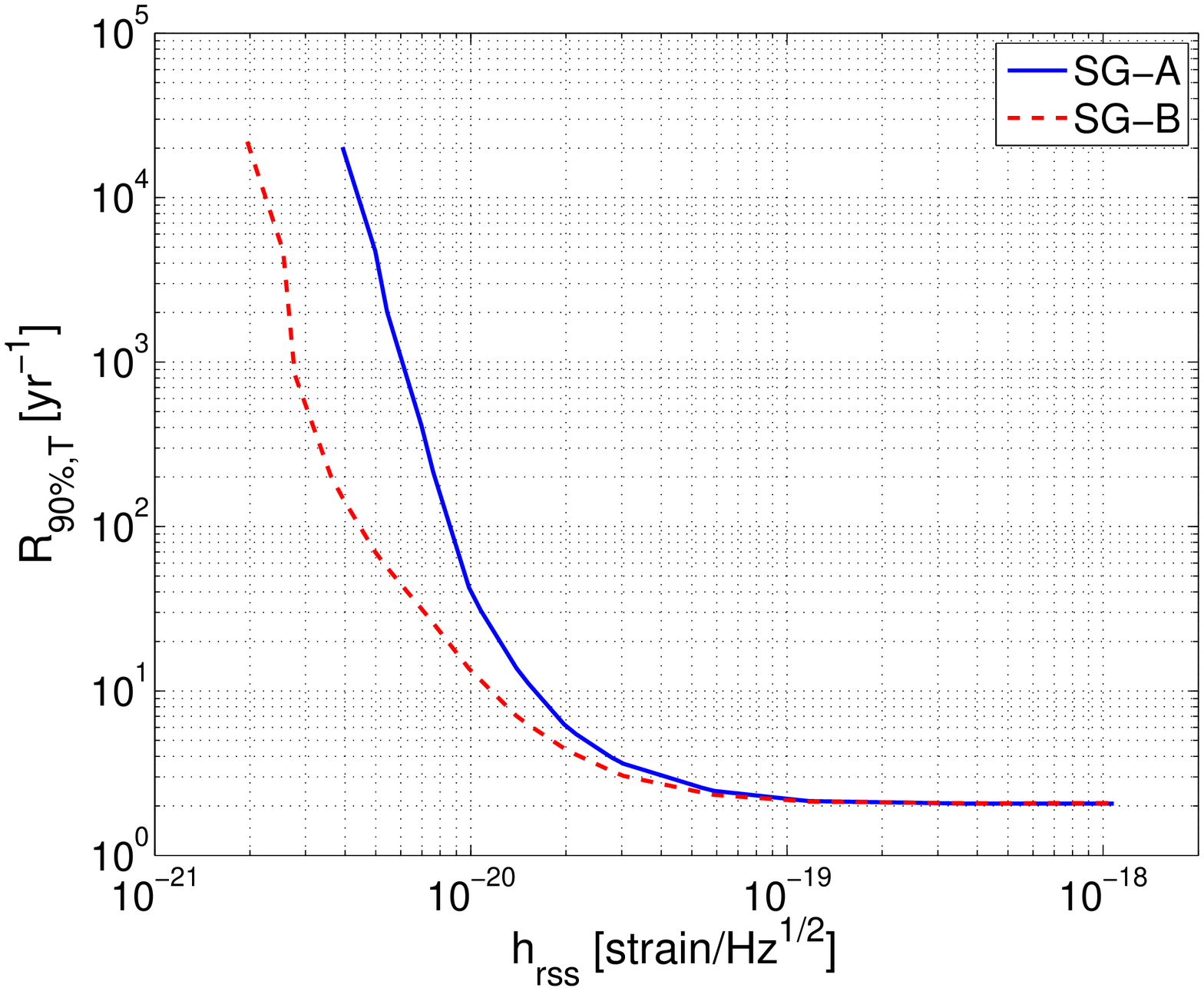}} \qquad\qquad
  \subfloat[]{\includegraphics[height=2.2in]{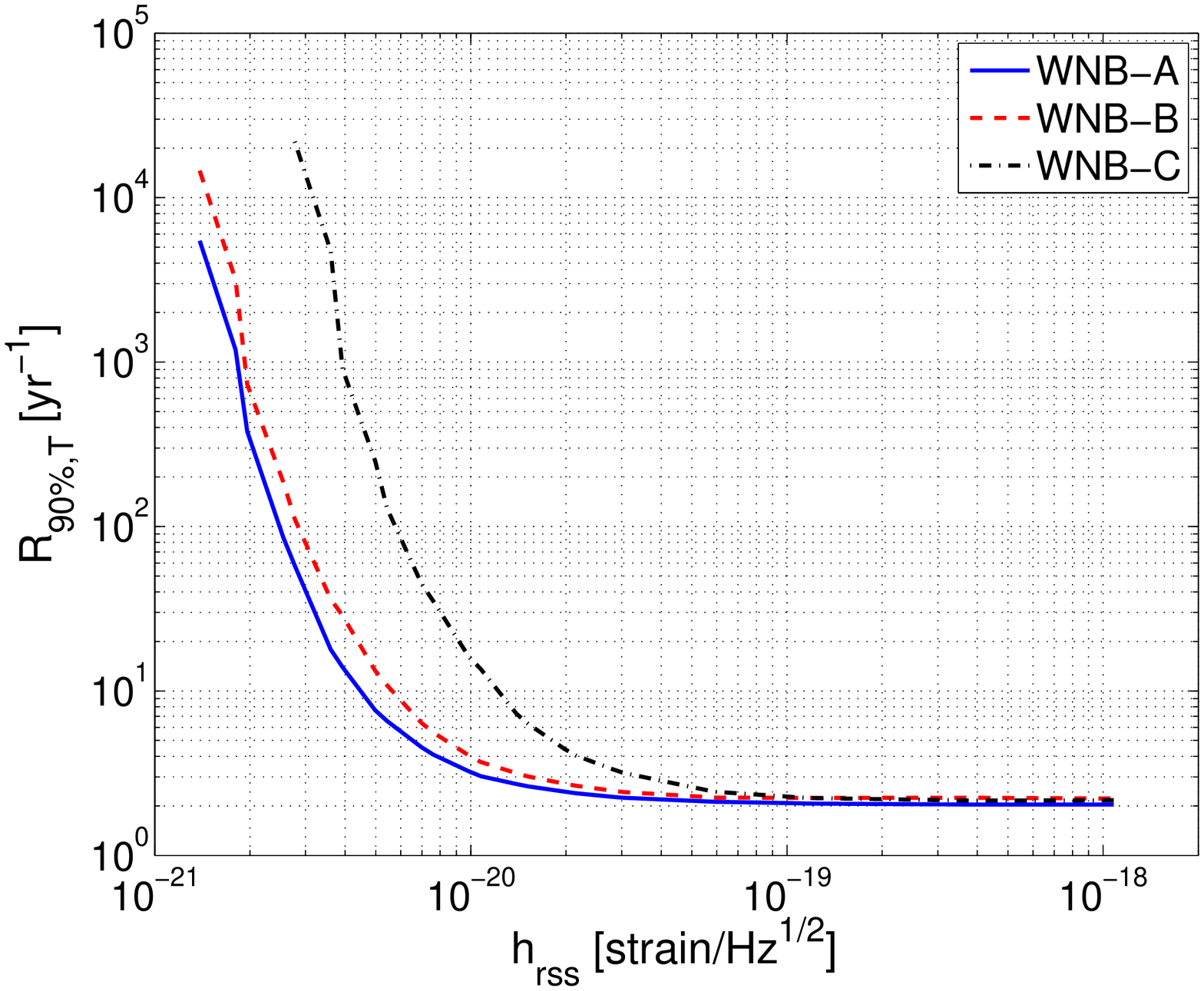}} \\
  \caption{Loudest event statistic upper limits for the 15 simulated GW signals used to test the sensitivity of the search (calculated with Equation~\ref{eq:LES_eff}). 1 $\sigma$ uncertainties are included by adjusting the signal amplitudes upward.
    Top left: ADI waveforms.
    Top right: sinusoidal waveforms.
    Bottom left: sine-Gaussian waveforms.
    Bottom right: white noise burst waveforms.}
  \label{fig:LES_upper_limits}
\end{figure*}

For signal models without a physical distance calibration, we use the loudest event statistic to calculate rate upper limits at 90\% confidence based on the search pipeline's efficiency:
\begin{equation}\label{eq:LES_eff}
R_{90\%,\text{T}} = \frac{2.3}{\nicesum_k \epsilon_{k} (\text{SNR}_{\Gamma,k}^\star) T_{\text{obs},k}}.
\end{equation}
Here, $\epsilon_k$ is the efficiency of search $k$ at a detection threshold $\text{SNR}_{\Gamma,k}^\star$ corresponding to the loudest zero-lag trigger from search $k$.
The resulting upper limits are in terms of number of observations per time, as indicated by the subscript $\text{T}$.
They are a function of signal strength in the form of root-sum-squared strain, $h_\text{rss}$, and are presented in Figure~\ref{fig:LES_upper_limits}.

Again, systematic uncertainty in the form of amplitude calibration uncertainty is the main source of uncertainty for these upper limits.
This uncertainty is accounted for by adjusting the signal amplitudes used in the sensitivity study (and shown in the efficiency curves in Figure~\ref{fig:efficiency}) upward by a multiplicative factor corresponding to the respective 1 $\sigma$ amplitude calibration uncertainty; this results in conservative upper limits.

\begin{figure}[hbtp!]
  \centering
  \includegraphics[height=2.2in]{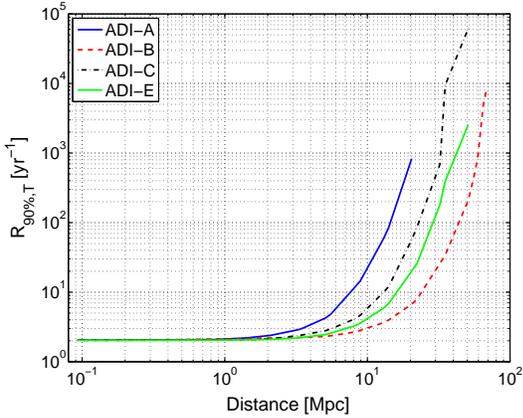}
  \caption{Loudest event statistic upper limits for the four ADI waveforms used to test the sensitivity of the search. Here, the upper limits are plotted in terms of distance rather than $h_\text{rss}$.}
  \label{fig:LES_upper_limits_ADI_dist}
\end{figure}

In Figure~\ref{fig:LES_upper_limits_ADI_dist}, we show the $R_{90\%,\text{T}}$ upper limits in terms of source distance for the ADI waveforms.
To compare these upper limits to the $R_{90\%,\text{VT}}$ upper limits, one would integrate the inverse of the $R_{90\%,\text{T}}$ curves (shown in Figure~\ref{fig:LES_upper_limits_ADI_dist}) over volume to obtain an overall estimate of the signal rate.
The two methods have been compared and the results are consistent within $\approx$25\% for all four of the ADI waveforms.
Differences between the two methods arise from the usage of different trigger ranking statistics (FAD vs. SNR$_\Gamma$), and the fact that the uncertainties are handled differently in each case.

\subsection{Discussion}
Given the absence of detection candidates in the search, we have reported upper limits on the event rate for different GW signal families.
Specifically, Figure~\ref{fig:LES_upper_limits} shows that, along with signal morphology, both the frequency and the duration of a signal influence the search sensitivity.
The $h_\text{rss}$ values for a search efficiency of 50\% obtained with S5 and S6 data are reported in Table~\ref{tab:results}. For the ADI waveforms, the 50\% efficiency distance is also given.
Although these limits cannot be precisely compared to the results of unmodeled short transient GW searches~\cite{Abadie:2012rq} because different waveforms were used by the two searches, it is clear that in order for long transient GW signals to be observed, it is necessary for the source to be more energetic: the total energy radiated is spread over hundreds of seconds instead of a few hundred milliseconds.

\begin{table*}[t]
  \centering
  \begin{tabular}{ c c c c c c c }
    \hline\hline
    Run      & \multicolumn{3}{c}{S5} & \multicolumn{3}{c}{S6}\\
    \hline
    Waveform & $h_\text{rss}^{50\%}$ [$\mathrm{Hz^{-1/2}}$] & $\mathrm{distance^{50\%}}$ [Mpc] & $E_\text{GW}$ [\msuncd] & $h_\text{rss}^{50\%}$ [$\mathrm{Hz^{-1/2}}$] & $\mathrm{distance^{50\%}}$ [Mpc] & $E_\text{GW}$ [\msuncd] \\
    \hline

    ADI-A  & $1.8 \times 10^{-21}$ & 5.4   & $1.5 \times 10^{-7}$ & $1.4 \times 10^{-21}$ &  6.8 & $9.4 \times 10^{-8}$ \\\hline
    ADI-B  & $1.9 \times 10^{-21}$ & 16.3  & $1.9 \times 10^{-7}$ & $1.7 \times 10^{-21}$ & 18.6 & $1.5 \times 10^{-7}$ \\\hline
    ADI-C  & $3.6 \times 10^{-21}$ & 8.9   & $9.9 \times 10^{-7}$ & $2.9 \times 10^{-21}$ & 11.3 & $6.3 \times 10^{-7}$ \\\hline
    ADI-E  & $2.3 \times 10^{-21}$ & 11.5  & $3.3 \times 10^{-7}$ & $2.0 \times 10^{-21}$ & 13.4 & $2.4 \times 10^{-7}$ \\\hline

    LINE-A & $3.9 \times 10^{-21}$ & -     & $4.9 \times 10^{-7}$ & $3.1 \times 10^{-21}$ & -    & $3.1 \times 10^{-7}$ \\\hline
    LINE-B & $8.5 \times 10^{-21}$ & -     & $9.6 \times 10^{-5}$ & $5.2 \times 10^{-21}$ & -    & $3.7 \times 10^{-5}$ \\\hline
    MONO-A & $1.3 \times 10^{-20}$ & -     & $3.1 \times 10^{-6}$ & $1.2 \times 10^{-20}$ & -    & $2.4 \times 10^{-6}$ \\\hline
    MONO-B & $2.1 \times 10^{-20}$ & -     & $2.3 \times 10^{-4}$ & $1.1 \times 10^{-20}$ & -    & $7.0 \times 10^{-5}$ \\\hline
    QUAD-A & $2.7 \times 10^{-21}$ & -     & $2.4 \times 10^{-7}$ & $2.4 \times 10^{-21}$ & -    & $1.9 \times 10^{-7}$ \\\hline
    QUAD-B & $5.1 \times 10^{-21}$ & -     & $1.6 \times 10^{-5}$ & $2.9 \times 10^{-21}$ & -    & $5.3 \times 10^{-6}$ \\\hline
    SG-A   & $2.4 \times 10^{-20}$ & -     & $1.0 \times 10^{-5}$ & $1.9 \times 10^{-20}$ & -    & $6.2 \times 10^{-6}$ \\\hline
    SG-B   & $2.1 \times 10^{-20}$ & -     & $2.5 \times 10^{-4}$ & $1.0 \times 10^{-20}$ & -    & $5.9 \times 10^{-5}$ \\\hline
    WNB-A  & $7.2 \times 10^{-21}$ & -     & $5.5 \times 10^{-6}$ & $5.5 \times 10^{-21}$ & -    & $3.2 \times 10^{-6}$ \\\hline
    WNB-B  & $9.1 \times 10^{-21}$ & -     & $1.8 \times 10^{-5}$ & $6.4 \times 10^{-21}$ & -    & $9.2 \times 10^{-6}$ \\\hline
    WNB-C  & $2.1 \times 10^{-20}$ & -     & $5.0 \times 10^{-4}$ & $1.3 \times 10^{-20}$ & -    & $1.7 \times 10^{-4}$ \\\hline

  \end{tabular}
  \caption{Values of $h_\text{rss}$ and distance where the search achieves 50\% efficiency for each of the simulated GW signals studied and each of the two datasets. $E_\text{GW}$ is an estimate of the energy released (given by Equation~\ref{eq:quadrupole}) by a source located at the detection distance for the ADI waveforms, or at \unit[10]{kpc} for the {\it ad hoc} waveforms.}
  \label{tab:results}
\end{table*}

One can also estimate the amount of energy emitted by a source located at a distance $r$ where the search efficiency drops below 50\% ($h_\text{rss}^{50\%}$) using the quadrupolar radiation approximation to estimate the energy radiated by a pair of rotating point masses:
\begin{equation}\label{eq:quadrupole}
E_\text{GW} \approx (h_\text{rss}^{50\%})^2 r^2 \pi^2 f_\text{GW}^2 \frac{c^3}{G}.
\end{equation} 

Considering the mean frequency of each GW waveform, we obtain an estimate of the energy that would have been released by a source that would be detected by this search.
For the ADI waveforms, the corresponding energy is between $9 \times 10^{-8}$ \msuncd\, and $6 \times 10^{-7}$ \msuncd.
For the {\it ad hoc} waveforms, one must fix a fiducial distance at which one expects to observe a signal.
For instance, considering a Galactic source at \unit[10]{kpc}, the emitted energy would be in the range $2\times 10^{-7}$--$2 \times 10^{-4}$ \msuncd.
This is still 2--4 orders of magnitude larger than the amount of energy estimated in~\cite{Mueller:2003fs} for a 10-kpc protoneutron star developing matter convection over \unit[30]{s}: $4 \times 10^{-9}$\, \msuncd.

Finally, we note that the search for long-duration transient signals is also closely related to the effort by LIGO and Virgo to observe a stochastic background of GWs.
One or more long-lived transient GW events, with a duration of days or longer, could produce an apparent signal in either the isotropic~\cite{Abbott:2011rs,Aasi:2014zwg,Aasi:2014jkh} or directional~\cite{Abbott:2011rr} stochastic GW searches.
It was for this reason that this long-duration transient detection pipeline was originally developed~\cite{Thrane:2010ri}.
The methods for detecting these long-duration transients have been adapted, in the study described in this present paper, to search for signals in the \unit[10]{s} to \unit[500]{s} regime.
A dedicated search for long-duration transient GW signals which last for days or longer will be a necessary component in the effort to understand the origin of apparent stochastic background signals which may be observed by LIGO and Virgo in the future~\cite{Thrane:2015wla}.

\section{Conclusions}\label{sec:conclusion}

In this paper, we have presented an all-sky search for long-lasting GW transients with durations between a few seconds and a few hundred seconds.
We performed the search on data from the LIGO H1 and L1 detectors collected during the S5 and S6 science runs.
We used a cross-correlation pipeline to analyze the data and identify potential GW candidate triggers.
To reject high-$\text{SNR}_\Gamma$ triggers due to detector noise, we defined a discriminant cut based on the trigger morphology.
We have also used data quality flags that veto well-identified instrumental or environmental noise sources to remove significant outliers.
No GW candidates were identified in this search, and as a consequence, we set upper limits on several types of simulated GW signals.
These are the first upper limits from an unmodeled all-sky search for long-transient GWs.
The upper limits are given in Table~\ref{tab:ADI_FAD_UL} and Figures~\ref{fig:LES_upper_limits} and~\ref{fig:LES_upper_limits_ADI_dist}.

After 2010, the LIGO and Virgo interferometers went through a series of upgrades~\cite{Aasi:2014jea,Acernese:2014hva}.
LIGO has just started its first observational campaign with its advanced configuration and will be joined by Virgo in 2016~\cite{Aasi:2013wya}.
The strain sensitivity of the advanced detectors is expected to eventually reach a factor of 10 better than the first-generation detectors.
This development alone should increase the distance reach of our search by a factor of 10, the energy sensitivity by a factor of 100, and the volume of space which we can probe by a factor of 1000.

Improvements are also being made to this search pipeline; a technique for generating triggers called ``seedless clustering'' has been shown to increase the sensitivity of the search by 50\% or more in terms of distance~\cite{Thrane:2013bea,Thrane:2014bma,Coughlin:2014xqa,Coughlin:2014swa,Coughlin:2015jka,Thrane:2015psa}.
The improvements to the search pipeline described in this paper, coupled with the increased sensitivity of LIGO and Virgo, will drastically improve the probability of detecting long-duration transient GWs and pave the way for an exciting future in GW astronomy.

\begin{appendix}

\section{Visible volume and false alarm density}\label{sec:vvis_FAD}

In order to constrain the rate and source density of the GW signals studied in this search, we estimate the volume of the sky in which the search algorithm is sensitive to these signals.
For this, we use the visible volume~\cite{Abadie:2011kd,Abadie:2012prd}
\begin{equation}\label{eq:visible_volume}
  V_\text{vis}(\text{SNR}_\Gamma) = \sum_i 4\pi r_i^2 \left( \frac{dN_\text{inj}}{dr} (r_i) \right) ^{-1}.
\end{equation}
Here the index $i$ runs over detected injections, $r_i$ is the distance to the $i$th injection, and $dN_\text{inj}/dr$ is the radial density of injections.
The SNR$_\Gamma$ parameter sets the threshold which determines whether an injection is recovered or not.
To calculate the visible volume, we require distance-calibrated waveforms, so this method is not practical for the {\it ad hoc} waveforms discussed previously.

Our estimate of the visible volume is affected by both statistical and systematic uncertainties.
We can estimate the statistical uncertainty on the visible volume using binomial statistics~\cite{Abadie:2011kd,Abadie:2012prd}:
\begin{equation}\label{eq:vvis_stat}
  \sigma_\text{stat} = \sqrt{\sum_i \left[ 4\pi r_i^2 \left( \frac{dN_\text{inj}}{dr} (r_i) \right) ^{-1} \right] ^2}
\end{equation}

Systematic uncertainty on the visible volume comes from the amplitude calibration uncertainties of the detectors:
\begin{equation}\label{eq:vvis_sys}
  \sigma_\text{sys} = 3 \times V_\text{vis} \times \sigma_\text{calibration} .
\end{equation}
These uncertainties are discussed further in Section~\ref{subsec:rates}.
We can estimate the total uncertainty on the visible volume by summing the statistical and systematic uncertainties in quadrature:
\begin{equation}\label{eq:vvis_toterr}
\sigma_{V_\text{vis}} = \sqrt{\sigma_\text{stat}^2 + \sigma_\text{sys}^2}.
\end{equation}
We note that the statistical uncertainty is negligible for this search compared to the systematic uncertainty from the amplitude calibration uncertainty of the detectors.

The false alarm density (FAD) statistic is useful for comparing the results of searches over different datasets or even using different detector networks~\cite{Abadie:2011kd,Abadie:2012prd,cWB:FAD}.
It provides an estimate of the number of background triggers expected given the visible volume and background livetime of the search.
The classical FAD is defined in terms of the FAR divided by the visible volume:
\begin{equation}\label{eq:FAD_classical}
\text{FAD}_\text{c} (\text{SNR}_\Gamma) = \frac{\text{FAR} (\text{SNR}_\Gamma)}{V_\text{vis} (\text{SNR}_{\Gamma})}.
\end{equation}
In this way, the FAD accounts for the network sensitivity to GW sources as well as the detector noise level and the accumulated livetime.

We follow~\cite{cWB:FAD} to define a FAD which produces a monotonic ranking of triggers:
\begin{equation}\label{eq:FAD_mono}
  \text{FAD} (\text{SNR}_{\Gamma,i}) = \text{min}(\text{FAD}_\text{c} (\text{SNR}_{\Gamma,i}), \text{FAD}_\text{c} (\text{SNR}_{\Gamma,i-1}) ),
\end{equation}
where the index $i$ runs over triggers in increasing order of $\text{SNR}_\Gamma$.

One then uses the FAD to combine results from searches over different datasets or with different detector networks by calculating the time-volume productivity of the combined search:
\begin{equation}\label{eq:tv_product}
\nu (\text{FAD}) = \nicesum_k V_{\text{vis},k} (\text{FAD}) \times T_{\text{obs},k}.
\end{equation}
Here the index $k$ runs over datasets or detector networks.
We note that the denominator in Equation~\ref{eq:LES_vvis} is equal to $\nu$ as described here.
The uncertainty on $\nu$ can be calculated as
\begin{equation}\label{eq:tv_sigma}
\sigma_\nu (\text{FAD}) = \sqrt{\nicesum_k T_{\text{obs},k}^2 \sigma_{V_\text{vis},k}^2 (\text{FAD})}
\end{equation}
The combined time-volume product is then used to calculate final upper limits (see Equations~\ref{eq:LES_vvis} and~\ref{eq:rate}).

\section{Bayesian formalism for FAD upper limit calculation}\label{sec:bayesian_FAD}
The observed astrophysical rate of triggers is given by
\begin{equation}\label{eq:rate}
R = \frac{N}{\nu (\text{FAD}^\star)},
\end{equation}
where $N$ is the number of triggers observed by the search, $\nu$ is the time-volume product described in Equation~\ref{eq:tv_product}, and FAD$^\star$ is the false alarm density of the most significant zero-lag trigger.
Although this search found no GW candidates, $N$ is Poisson distributed, with some underlying mean and variance $\mu$.
To properly handle the uncertainty associated with the time-volume product, we utilize a Bayesian formalism to account for uncertainties in both quantities in the calculation of rate upper limits.
Here, our goal is to constrain the rate using the mean expected number of triggers, $\mu$.

Given an observation of $N$ triggers with a time-volume product of $\overline{\nu}$, we can define a posterior distribution for $\mu$ and $\nu$ in terms of Bayes' theorem.
\begin{equation}\label{eq:post}
P(\mu,\nu | N,\overline{\nu}) = \frac{P(N, \overline{\nu} | \mu, \nu) P(\mu,\nu)}{P(N, \overline{\nu})},
\end{equation}
where $P(\mu,\nu | N,\overline{\nu})$ is the posterior distribution, $P(N, \overline{\nu} | \mu, \nu)$ is the likelihood, $P(\mu,\nu)$ are the priors, and $P(N, \overline{\nu})$ is the evidence.

We may disregard the evidence since it is simply a normalization factor, and use uniform priors on $\mu$ and $\nu$.
\begin{subequations}\label{eq:priors}
  \begin{align}
    \mu &\in [0,\mu_\text{max}],\\
    \nu &\in [0,\nu_\text{max}].
  \end{align}
\end{subequations}
Here $\mu_\text{max}$ and $\nu_\text{max}$ are defined to be large enough that all of the significant posterior mass is enclosed.

The likelihood function can be framed as the product of two separate distributions: a Poisson distribution with mean $\mu$ and a Gaussian distribution with mean $\overline{\nu}$ and sigma $\sigma_\nu$ (see Equation~\ref{eq:tv_sigma}).
\begin{equation}\label{eq:likelihood}
P(N,\overline{\nu} | \mu,\nu) = \frac{e^{-\mu}\mu^N}{N!}\frac{e^{-(\nu-\overline{\nu})^2/(2\sigma_\nu^2)}}{\sqrt{2\pi}\sigma_\nu}.
\end{equation}

Using this formalism, we can calculate the posterior distribution as a function of $\mu$ and $\nu$.
However, we are primarily interested in setting limits on the rate, given as the ratio of $\mu$ and $\nu$.
We can transform the posterior distribution to be a function of two new variables by multiplying the original posterior by the Jacobian determinant of the transformation:
\begin{equation}\label{eq:transform}
P(R,\nu | N,\overline{\nu}) = P(\mu,\nu | N,\overline{\nu}) \times |J|.
\end{equation}
Choosing the new variables to be $R$ and $\nu$ (for simplicity), we obtain a Jacobian determinant of
\begin{equation}\label{eq:Jacobian}
|J| =
\begin{vmatrix}
\dfrac{\strut\partial \mu}{\strut\partial R} & \dfrac{\strut\partial \mu}{\strut\partial \nu} \\
\dfrac{\strut\partial \nu}{\strut\partial R} & \dfrac{\strut\partial \nu}{\strut\partial \nu}
\end{vmatrix}
= \nu.
\end{equation}
This gives a posterior distribution of
\begin{equation}\label{eq:postRnu}
P(R,\nu | N,\overline{\nu}) = \nu \frac{e^{-\mu}\mu^N}{N!}\frac{e^{-(\nu-\overline{\nu})^2/(2\sigma_\nu^2)}}{\sqrt{2\pi}\sigma_\nu}.
\end{equation}

Finally, we marginalize over $\nu$ to get the posterior distribution of $R$:
\begin{equation}\label{eq:Rpost1D}
P(R | N,\overline{\nu}) = \int_0^{\nu_\text{max}} P(R,\nu | N,\overline{\nu}) d\nu.
\end{equation}
Using this distribution, we can find the 90\% limit on $R$, $R_{90\%,\text{VT}}$, such that 90\% of the posterior mass is enclosed, by numerically solving Equation~\ref{eq:bayes_90}.

\begin{equation}\label{eq:bayes_90}
  0.9 = \int_0^{R_{90\%,\text{VT}}} P(R | N,\overline{\nu}) dR.
\end{equation}

\end{appendix}

\begin{acknowledgments}
The authors gratefully acknowledge the support of 
the United States National Science Foundation (NSF) for the construction and operation of the LIGO Laboratory,
the Science and Technology Facilities Council (STFC) of the United Kingdom, 
the Max-Planck-Society (MPS), and the State of Niedersachsen/Germany 
for support of the construction and operation of the GEO600 detector,
the Italian Istituto Nazionale di Fisica Nucleare (INFN) and 
the French Centre National de la Recherche Scientifique (CNRS)
for the construction and operation of the Virgo detector
and the creation and support  of the EGO consortium. 
The authors also gratefully acknowledge research support from these agencies as well as by 
the Australian Research Council,
the International Science Linkages program of the Commonwealth of Australia,
the Council of Scientific and Industrial Research of India, 
Department of Science and Technology, India,
Science \& Engineering Research Board (SERB), India,
Ministry of Human Resource Development, India,
the Spanish Ministerio de Econom\'ia y Competitividad,
the Conselleria d'Economia i Competitivitat and Conselleria d'Educació, Cultura i Universitats of the Govern de les Illes Balears,
the Foundation for Fundamental Research on Matter supported by the Netherlands Organisation for Scientific Research, 
the National Science Centre of Poland, 
the European Union,
the Royal Society, 
the Scottish Funding Council, 
the Scottish Universities Physics Alliance, 
the National Aeronautics and Space Administration, 
the Hungarian Scientific Research Fund (OTKA),
the Lyon Institute of Origins (LIO),
the National Research Foundation of Korea,
Industry Canada and the Province of Ontario through the Ministry of Economic Development and Innovation, 
the National Science and Engineering Research Council Canada,
the Brazilian Ministry of Science, Technology, and Innovation,
the Carnegie Trust, 
the Leverhulme Trust, 
the David and Lucile Packard Foundation, 
the Research Corporation, 
and the Alfred P. Sloan Foundation.
The authors gratefully acknowledge the support of the NSF, STFC, MPS, INFN, CNRS and the
State of Niedersachsen/Germany for provision of computational resources. 

\end{acknowledgments}

\bibliography{biblio}
\
\end{document}